\documentclass[manuscript]{aastex}

\usepackage{natbib}
\usepackage{epsfig}
\usepackage[update,prepend]{epstopdf}
\usepackage{morefloats}

\shorttitle{TTV}
\shortauthors{Mazeh et al.}

\begin{document}

\title{TRANSIT TIMING OBSERVATIONS FROM {\it KEPLER}. VIII
CATALOG OF
TRANSIT TIMING MEASUREMENTS OF THE FIRST TWELVE QUARTERS}


\author{Tsevi Mazeh\altaffilmark{1},
Gil Nachmani\altaffilmark{1},
Tomer Holczer\altaffilmark{1},
Daniel C. Fabrycky\altaffilmark{2},
Eric B. Ford\altaffilmark{3},
Roberto Sanchis-Ojeda\altaffilmark{4},
Gil Sokol\altaffilmark{1},
Jason F. Rowe\altaffilmark{5},
Shay Zucker\altaffilmark{6},
Eric Agol\altaffilmark{7},
Joshua A. Carter\altaffilmark{8},
Jack J. Lissauer\altaffilmark{5},
Elisa V. Quintana\altaffilmark{9},
Darin Ragozzine\altaffilmark{3},
Jason H. Steffen\altaffilmark{10},
William Welsh\altaffilmark{11}
}

\altaffiltext{1}{ School of Physics and Astronomy, Raymond and
Beverly Sackler Faculty of Exact Sciences, Tel Aviv University,
Tel Aviv 69978, Israel}
\altaffiltext{2}{Department of Astronomy and Astrophysics,
University of Chicago, 5640 Ellis Ave., Chicago, IL 60637, USA}

\altaffiltext{3}{Astronomy Department, University of Florida,
Gainesville, FL 32111, USA}

\altaffiltext{4}{Department of Physics, and Kavli Institute for
Astrophysics and Space Research, Massachusetts Institute of
Technology, Cambridge, MA 02139, USA}

\altaffiltext{5}{NASA Ames Research Center, Moffett Field, CA
94035, USA}

\altaffiltext{6}{Department of Geophysical, Atmospheric and
Planetary Sciences,
Raymond and Beverly Sackler Faculty of Exact Sciences Tel Aviv
University, 69978 Tel Aviv, Israel}
\altaffiltext{7}{Department of Astronomy, Box 351580, University
of Washington, Seattle, WA 98195, USA}

\altaffiltext{8}{Hubble Fellow, Harvard-Smithsonian Center for
Astrophysics, 60 Garden Street, Cambridge, MA 02138, USA}

\altaffiltext{9}{SETI Institute, 189 Bernardo Ave, Suite 100,
Mountain View, CA 94043, USA}

\altaffiltext{10}{Fermilab Center for Particle Astrophysics, P.O.
Box 500, MS 127, Batavia, IL 60510, USA}

\altaffiltext{11}{Astronomy Department, San Diego State
University,
5500 Campanile Drive, San Diego, California 92182, USA}

\begin{abstract}

Following \citet{ford11, ford12b} and \citet{steffen12b} we
derived the transit timing of $1960$ {\it Kepler} KOIs using the
pre-search data conditioning (PDC) light curves of the first twelve quarters
of the {\it Kepler} data.
For $721$ KOIs with large enough SNRs, we obtained also the duration and depth of each
transit.
The results are presented as a catalog for the community to use.
We derived a few statistics of our results that could be used to
indicate significant variations.
Including systems found by previous works, we have found
$130$ KOIs that showed highly significant TTVs, and
$13$ that had short-period TTV modulations with small amplitudes.
We consider two effects that could cause {\it apparent}
periodic TTV --- the finite sampling of the observations and 
the interference with the stellar activity, stellar spots in particular.
We briefly discuss some statistical aspects of our detected
TTVs. We show that the TTV period is correlated with the orbital
period of the planet and with the TTV amplitude.
\end{abstract}


\keywords{planetary systems –-- planets and satellites:
detection –--
techniques: miscellaneous --- technique: photometric}

\section{Introduction}
Since 2009 May 2, the {\it Kepler} spacecraft has been collecting
science-quality photometric data of more than 150,000 stars.
Based on the first 5 months of data, \citet[][hereafter
B11]{borucki11}
identified 1235 planet candidates associated with 997 host stars.
Analysis of the first 16 months of data \citep[][hereafter
B12]{batalha12} yielded additional 1091 viable planet
candidates --- termed Kepler objects of interest, or KOIs. The
almost uninterrupted accurate {\it Kepler} light
curves of these KOIs enable the community to detect minute
changes in the observed transit light curves.  This is especially
true for the individual times of transit, which for some KOIs
show variation (=TTV) relative to a linear ephemeris that
assumes a constant Keplerian orbit. These TTVs can indicate a
dynamical interaction with additional objects in the system, as
was predicted by the seminal works of  \citet{holman05} and
\citet{agol05}. Indeed, TTVs turn out to be a crucial tool
in the study of systems with known multiple transiting planets
\citep[e.g.,][]{holman10, lissauer11a, cochran11, fabrycky12,
steffen12a,lithwick12}.

However, TTVs can do much more. They may indicate dynamical
interactions with unseen, otherwise undetected, additional
objects in the
system \citep[e.g.,][]{ballard11,nesvorny12,nesvorny13}.
Therefore, it can be useful to
perform a systematic search for TTV in all KOIs, as was done in
the work
of \citet[][hereafter F11]{ford11} and was continued with
\citet[][hereafter
F12]{ford12b} and
\citet[][hereafter S12]{steffen12b}
works, based on the first six quarters of Kepler data.
This paper is a follow-up of F11, F12 and S12 studies (see also
the catalog of Rowe
et al., private communication), presenting a systematic analysis
of the
first {\it twelve} quarters of the {\it Kepler} data of all KOIs.
The goal
is to produce an easy-to-use catalog that can stimulate
further analysis of interesting systems and statistical analysis
of the sample of Kepler KOIs with significant TTVs.

After presenting the details of our pipeline and the catalog
itself (Section~\ref{analysis}), we  derive a few statistical
characteristics of each TTV series that can identify the ones
with significant variations (Section 3). Sections 4 and 5 list
$143$
systems with highly significant TTVs, and Section~\ref{comments}
comments on some interesting systems, in particular the ones for
which the derived TTVs could be of a non-dynamical origin.
In Section~\ref{discussion} we present a few basic statistical
features of the sample of the
$130$
systems, and briefly discuss the possible use of the catalog.

\section{Analysis of the transit light curves}
\label{analysis}

The catalogs of B11 and B12
(http://archive.stsci.edu/kepler/planet\_ candidates.html)
listed $2321$ KOIs, out of which we did not analyze $21$ KOIs.
These included $13$ 
KOIs for which B12 had only one transit
(B12 had 20 systems with only one transit, but since then Kepler additional data showed more transits for seven of them, so we were left with only 13 with one transit), 
one KOI that did not have a measured
transit duration, and seven KOIs with transit duration of less
than one
hour, too short for our analysis.
We therefore analyzed the light curves of $2300$
KOIs. We used the publicly available PDC long-cadence
(ftp://archive.stsci.edu/pub/kepler/lightcurves/tarfiles) data, which used the BJD$_{TDB}$ timings.

We started by phase-folding the {\it Kepler} light curve of each KOI
with its ephemeris in B12, in order to obtain the best
possible template for the transit light curve (see below for
details). We used the best-fit transit model as a template to
measure the actual timing of each individual transit timing (=TT)  and derive
its O-C --- the difference between the TT and the expected time,
based on
the linear ephemeris. For KOIs with high enough SNR (see below),
we derived the TTs while allowing the duration and depth of each
transit to vary as well.  Considering our template just as a
mathematical function, finding the timing, duration and depth of a transit
was equivalent to moving the
center of the template or stretching it in the time and flux
dimensions. In our approach, we searched for the minimum of the
sum-of-squared residuals, the standard $\chi^2$ function, in the
three-parameter space. Similarly to \citet{ford11},  we iterated the procedure,
aligning the transits based on their measured timings, in order
to generate a better transit model, and then re-analyzed the
individual transits.
Finally, we modified $T_0$, the timing of the first transit, and
the period of each KOI, whenever we detected a significant linear
trend in our O-Cs, and re-derived the O-Cs relative to the new
ephemerides.

\begin{figure}[h]
\centering
\resizebox{12cm}{10cm}
{\includegraphics{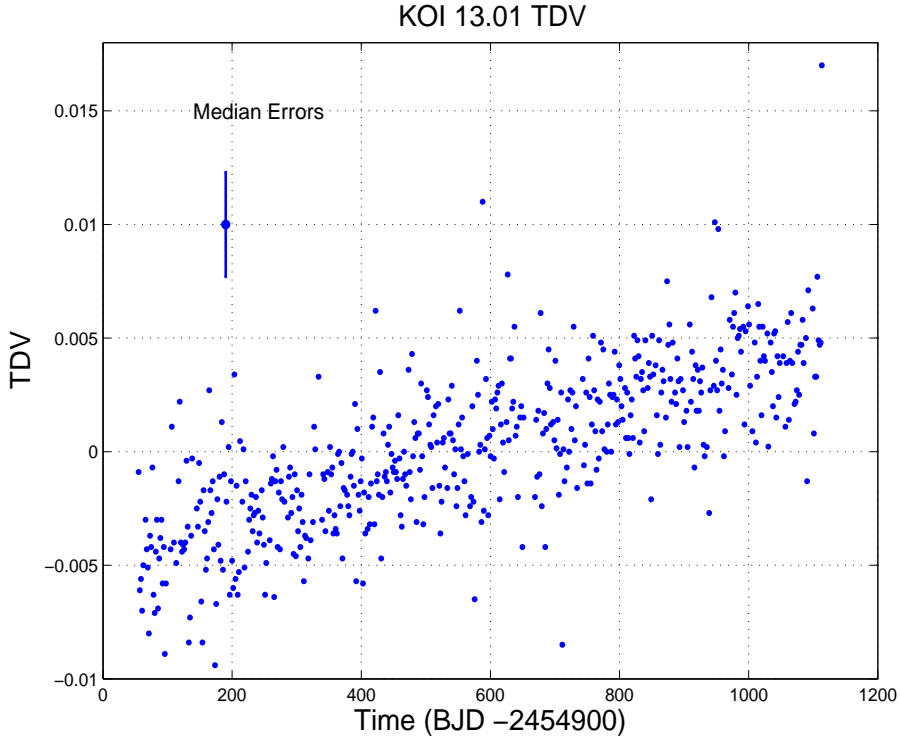}}
\caption{The TDV (duration variation) of KOI-13.01. Each point
represents our best estimate for the deviation of the transit
duration from its averaged value, in units of the averaged duration.
A typical error is included in the figure.
}
\label{KOI13_TDV}
\end{figure}

Although the main focus of this paper was the TTVs of the
KOIs, we opted to vary the three parameters of the template
simultaneously for KOIs with high enough SNRs,
because we found a few KOIs for which the transit
duration or depth did vary significantly, either because of
physical processes, or as a result of some observational
effects.
One example is KOI-13, for which \citet{szabo12mnras}
have
found some indications for a long-term variation of the impact
parameter, equivalent to detecting variation of the transit
duration. Our analysis, now based on twelve quarters, confirms
the result of 
\citet{szabo12mnras}, and is presented in
Figure~\ref{KOI13_TDV}. One can see the highly
significant linear duration variation of KOI-13, which amounts to
$\sim1\%$ peak-to-peak modulation over the entire data span. 
For such cases the simultaneous analysis of timing,
duration and depth
is an advantage, and, in principle, can yield better timing of
each transit.

However, for low-SNR transits, minimizing the $\chi^2$
function with respect to the three parameters
simultaneously could yield a completely erroneous result, based on
some accidental local minimum in a noisy $\chi^2$ surface. In
fact, for systems with even lower SNR, fitting the timing alone
could yield misleading minima. We
therefore divided the KOIs into three groups, according to their typical
SNR for a {\it single} transit, defined as:

\begin{equation}
{\rm SNR} = \frac{\delta}{\sigma}\sqrt{N} \ ,
\end{equation}
where
$\delta$ is the relative transit depth,
$\sigma$ is the relative uncertainty per point, derived from the
scatter of the light curve outside the transit,
and
$N$ is the averaged number of points per single transit.
For each KOI we derived the median of its transit SNRs.

We considered three classes of KOIs:
\begin{itemize}

\item
${\rm SNR} < 2.5$ --- This class included $340$ KOIs. We did not perform any analysis for objects of this class, due to the poor SNR. 
\item
$2.5 < {\rm SNR} < 10$ --- This class included $1239$ KOIs, for which we have calculated TTs, while fixing the transit duration and depth, derived from the best-fit model. 
\item
$SNR >10$  --- For the 721 KOIs in this class, we  derived simultaneously the transit timing, duration and depth.  
\end{itemize}

\subsection{Transit model}

Our default choice for the transit templates was the
\citet{mandel02} model, which we derived for each KOI's folded
light curve through a $\chi^2$ minimization. However, since some transits showed slight
asymmetries, e.g., KOI-13 \citep[][]{szabo11, mazeh12}, 
 and other transits had SNR values which were too low
for a convincing Mandel-Agol fit, we used two additional models
as possible templates: "Legendre-based" and "Fermi-based" models,
which are described below.  
We computed these three models for each KOI, and chose the model with the lowest
$\chi^2$ value
as the transit template. However, due to the
astrophysical basis of the Mandel-Agol model --- in contrast to the
other two which were merely mathematical heuristics --- we preferred
the Mandel-Agol model whenever it gave a good enough fit.
Hence, we chose the Mandel-Agol model also in cases where its
r.m.s.~exceeded the r.m.s.~of the other two models by up to 5\%.

Below we provide a few details about the three models:

\begin{itemize}
\item
We used our own code to fit the \citet{mandel02} model, with a
quadratic limb-darkening law, using coefficients that we interpolated from \citet{claret11} tables,
assuming  $log \ g=4$, Solar metallicity, and {\it Kepler} KIC temperature \citep{brown11}.
 
\item
The Legendre-based model had the form:

\begin{equation}
F\left (\tau \right ) =
\sum_{k=1}^{N_1} A_kL^k\left (\tau \right )
+ \sum_{k=1}^{N_2} S_k\, sin\left (\pi k\tau \right)
+ \sum_{k=1}^{N_3} C_k\, cos\left (\frac{\pi}{2} k \tau \right)
\,,
\end{equation}
where $L^{k}$ was the Legendre polynomial of order $k$, $\tau$
was
the normalized phase of the transit, such that at the beginning
of ingress $\tau\left ( t_{1} \right )=-1$, and at the end of
 egress $\tau\left ( t_{4} \right )=+1$,
$A_{k},S_{k}$ and $C_{k}$
were linear parameters found analytically,
and $N_1 = N_2 = N_3 = 10$ were the maximum orders we allowed for
each function. We optimized the model by varying the phases
of $t_1$ and $t_4$ within the orbital period. We avoided local
bumps in the model by reducing its three orders ($N_1$, $N_2$ and
$N_3$) separately and by using linear fits to overcome local
changes in convexity.

\item
The Fermi-based model had the form:

\begin{equation}
F\left ( \tau  \right )=
1 + M\left [
\frac{1}{e^{\left (\tau +\varphi +\mu   \right )/s}+1}
+
\frac{1}{e^{\left (\tau+\varphi  -\mu   \right )/s}+1}
-1
\right ],
\end{equation}
where $\tau$ was the phase of the transit, as for the Legendre
model, and $\varphi$, $s$, $\mu$ and $M$ were free parameters,
standing for the transit phase, ingress and egress steepness, width and depth of the
transit.
In order to obtain a more "transit-like" shape, we replaced the points
at the bottom part of the transit with a parabola, with its
width as another free parameter, under the constraint that
the resulting function and its first derivative were both
continuous. 
\end{itemize}

For most KOIs the pipeline selected the \citet{mandel02}
model (1829 KOIs). It chose the Legendre-based model when there was a
significant asymmetry in the folded light curve of the transit
(87 KOIs), and the Fermi-based one only when the SNR of
the folded light curve was low (44 KOIs).

\subsection{Finding the timing, duration and depth of each
transit and their uncertainties}

 We analyzed each transit after fitting a
polynomial to the light curve on the two sides of the transit,
in order to remove stellar and instrumental long-term
photometric variations during the transit.

We derived the timing, and when appropriate the duration and
depth, of each transit by
minimizing the standard
$\chi^2$ function using the \begin{footnotesize}MATLAB FMINSEARCH \end{footnotesize} function, based on the Nelder-Mead Simplex
method
\citep{lagarais98}, assuming each measurement had the same error.
Our pipeline then made sure that the $\chi^2$-minimum found was
indeed the global minimum by an automated grid search over the parameter space. 
We then used the $\mathcal{F}$-test to compare the transit model with
the timing (and duration and depth when appropriate) found against
a constant flux assumption (no transit at all), and rejected all transits with an $\mathcal{F}$-test
False Alarm Probability (=FAP) larger than 0.025. For these cases, 
the transit timing table quotes no timing (nor duration and depth).

We estimated the errors of the three quantities from the inverted 
Hessian matrix, calculated at the minimum. The error of each individual {\it Kepler} measurement
was based on the scatter of the light curve around the
polynomial fit before and after each transit. When the
Hessian matrix turned out to be singular, we assigned an error
that was equal to the median of the other errors derived for
the KOI in question. Whenever that was the case we marked
the error with an asterisk in the table of transit timings.

For each KOI, we ignored outlying timing, duration and depth values when their
corresponding O-C values were too different  from the other O-Cs of that KOI, or
 their error estimate was too large. Usually, a large error meant that
some photometric measurements during that transit were erroneous.
We rejected outliers based on both global and local mean and scatter. A value was considered an outlier if it deviated from the mean by more than five times the scatter of the series, defined as 1.4826 times its Median Absolute Deviation (MAD), plus three times its own error. 

\begin{figure}[h]
\centering
\resizebox{12cm}{10cm}
{\includegraphics{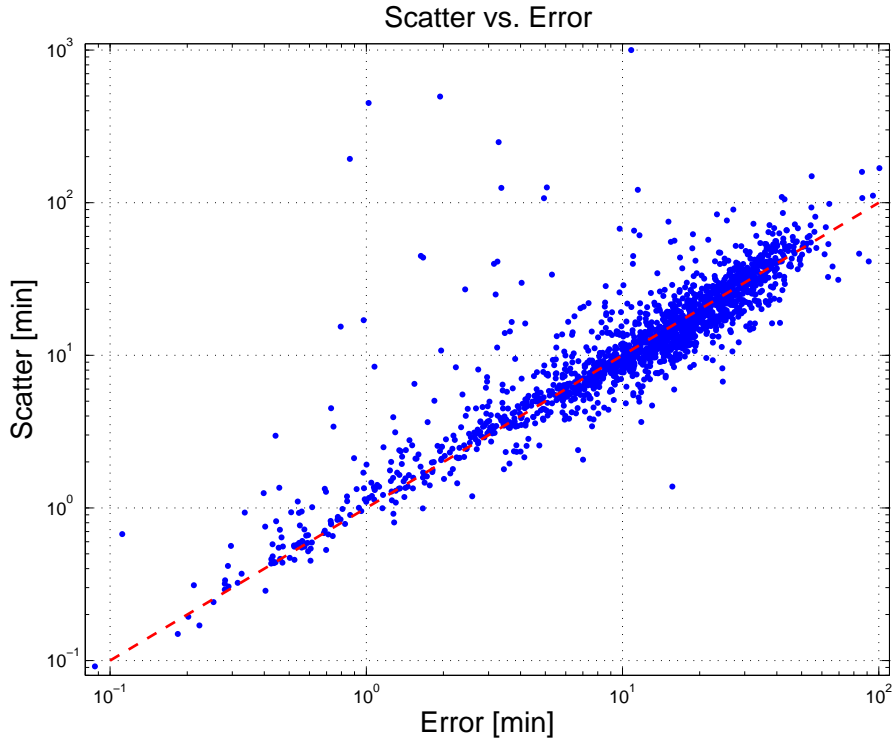}}
\caption{The scatter of the derived O-C timings as a function of
their typical uncertainty for all $1960$ KOIs.
}
\label{MADvsMed_error}
\end{figure}

In order to check the obtained uncertainties for the transit
timings, we computed for each KOI the scatter of its O-C values,
$s_{\rm O-C}$, and compared it with its typical error, derived
for each KOI by the median of its timing uncertainties ---
$\overline{\sigma}_{\rm TT}$. We expect these two values to be
similar for systems with no significant TTV. This is indeed the
case, as can be seen in
Figure~\ref{MADvsMed_error}.
The KOIs with O-C scatter larger than their uncertainties are those with significant TTVs.

\begin{figure}[h!]
\centering
\resizebox{12cm}{10cm}
{\includegraphics{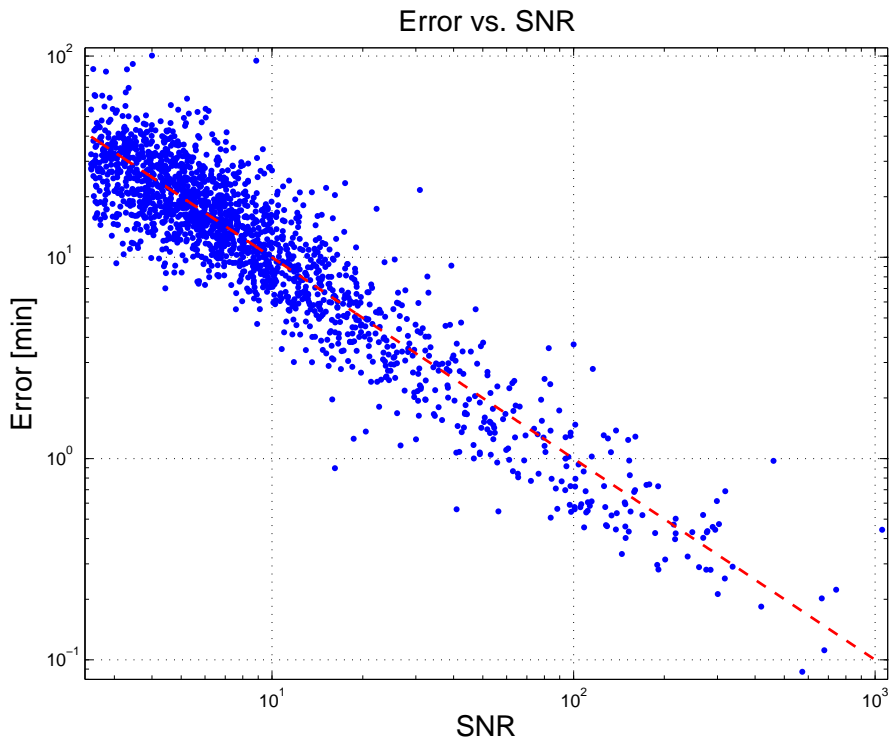}}
\caption{Typical transit timing uncertainty as a function of the 
 typical SNR of a single transit for each KOI.
The dashed red line represents $\overline{\sigma}_{\rm TT}=
100/$SNR }
\label{Med_errors_vsSNR}
\end{figure}

Another approach to check our error estimate is to compare the 
typical derived error of a KOI with the SNR of its transit. One can
expect the timing precision to improve with higher SNR.
In order to see whether this is really the case,
Figure~\ref{Med_errors_vsSNR} shows the median error, $\overline{\sigma}_{\rm TT}$, versus
the median SNR of that KOI, presenting a tight correlation
over the whole range of SNR, which goes from 2.5 to 1000. The plot
is consistent with the simple relation
\begin{equation}
\overline{\sigma}_{\rm TT} \simeq \frac{100} {{\rm SNR}} \ ,
\end{equation}
which is also plotted in the figure.
%

The last two figures suggest that our error estimate is
realistic.

\subsection{The catalog}
We present our results in two tables, available at
ftp://wise-ftp.tau.ac.il/pub/tauttv/TTV.
Table~\ref{tab:ephemeris} lists the modified ephemerides of the
KOIs, based on our analysis, together with the durations and depths of their
transits, derived from the folded light curve. The transit duration is quoted as a
fraction of the orbital period and the depth in units of the stellar
intensity outside the transit. Table~\ref{tab:TTV} lists our derived O-Cs,
relative to our modified ephemerides,
 for 167934 transits of 1960 KOIs with
${\rm SNR}> 2.5$. Of those, duration and depth changes, in
units of the transit model duration and depth, are given
for $62802$  transits of 721 KOIs with $ {\rm SNR} > 10$.

\begin{table}
\footnotesize \caption{Linear ephemerides of the KOI transits,
together with their durations and depths}
\begin{tabular}{cccccc}
\hline \hline
KOI     &  ~~~~~~T$_0$\tablenotemark{a} & ~~~~~~~~~~Period\tablenotemark{b} &
~Duration\tablenotemark{c} &
Depth\tablenotemark{d}&SNR\tablenotemark{e}\\
           &   ~~~~~~ [d]   & ~~~~~~~~~~[d] &  &  & \\
\end{tabular}

\begin{tabular}{r r r r r r }
\hline
$ 1.01 $ & $ 55.762538 $ & $ 2.47061337 $ & $ 0.0315 $ & $
0.01419 $ & $ 573.4 $ \\
   & $ \pm 0.000009 $ & $ \pm 0.00000004 $ &    &    &    \\
$ 2.01 $ & $ 54.357833 $ & $ 2.20473534 $ & $ 0.0764 $ & $
0.00669 $ & $ 317.3 $ \\
   & $ \pm 0.000019 $ & $ \pm 0.00000006 $ &    &    &    \\
$ 3.01 $ & $ 57.812640 $ & $ 4.88780191 $ & $ 0.0222 $ & $
0.00433 $ & $ 300.8 $ \\
   & $ \pm 0.000074 $ & $ \pm 0.00000058 $ &    &    &    \\
$ 4.01 $ & $ 90.526015 $ & $ 3.84937129 $ & $ 0.0298 $ & $
0.00132 $ & $ 31.1 $ \\
   & $ \pm 0.000315 $ & $ \pm 0.00000186 $ &    &    &    \\
$ 5.01 $ & $ 65.973089 $ & $ 4.78032914 $ & $ 0.0186 $ & $
0.00098 $ & $ 34.4 $ \\
   & $ \pm 0.000198 $ & $ \pm 0.00000144 $ &    &    &    \\
$ 7.01 $ & $ 56.611453 $ & $ 3.21366766 $ & $ 0.0552 $ & $
0.00074 $ & $ 24.1 $ \\
   & $ \pm 0.000359 $ & $ \pm 0.00000184 $ &    &    &    \\
$ 10.01 $ & $ 54.118640 $ & $ 3.52249863 $ & $ 0.0391 $ & $
0.00937 $ & $ 127.4 $ \\
   & $ \pm 0.000057 $ & $ \pm 0.00000031 $ &    &    &    \\
$ 12.01 $ & $ 79.595944 $ & $ 17.85521101 $ & $ 0.0172 $ & $
0.00917 $ & $ 318.6 $ \\
   & $ \pm 0.000413 $ & $ \pm 0.00001133 $ &    &    &    \\
$ 13.01 $ & $ 53.565019 $ & $ 1.76358759 $ & $ 0.0790 $ & $
0.00460 $ & $ 419.1 $ \\
   & $ \pm 0.000011 $ & $ \pm 0.00000003 $ &    &    &    \\
$ 17.01 $ & $ 54.485821 $ & $ 3.23469919 $ & $ 0.0477 $ & $
0.01078 $ & $ 239.0 $ \\
   & $ \pm 0.000034 $ & $ \pm 0.00000018 $ &    &    &    \\

\end{tabular}
\tablecomments{
$^a$T$_0$ in BJD -- 2454900. 
$^b$Orbital period. 
$^c$Transit duration in units of the orbital period. 
$^d$Transit depth in units of the stellar intensity outside the
transit. 
$^e$Median single-transit SNR.
\\
(This table is available in its entirety in a machine-readable
form
in ftp://wise-ftp.tau.ac.il/pub/tauttv/TTV. A portion is shown
here for
guidance
regarding its form and content.)}
\label{tab:ephemeris}
\end{table}

\begin{table}
\footnotesize \caption{O-C, duration (TDV) and depth (TPV)
changes of the transits}
\begin{tabular}{rr|rr|rr|rr}
\hline \hline
n\tablenotemark{a}  & t$_n\tablenotemark{b}$~~~& O-C$_n$\tablenotemark{c} & $\sigma_{n}$\tablenotemark{d}~&
~~TDV$_n$\tablenotemark{e}
&  $\sigma_{n}$\tablenotemark{f} ~& ~~TPV$_n$\tablenotemark{g} & $\sigma_{n}$\tablenotemark{h}~~\\
  &  [d]~~~ & [min]  & [min] & & & & \\

  \hline
$ 0 $ & $ 55.7625 $ & $ -0.057 $ & $ 0.085 $ & $ 0.0009 $ & $
0.003 $ & $ -0.0048 $ & $ 0.0028 $ \\
$ 1 $ & $ 58.2332 $ & $ 0.054 $ & $ 0.074 $ & $ -0.0015 $ & $
0.0023 $ & $ -0.0067 $ & $ 0.0023 $ \\
$ 2 $ & $ 60.7038 $ & $ -0.042 $ & $ 0.098 $ & $ 0.0019 $ & $
0.0028 $ & $ -0.01 $ & $ 0.003 $ \\
$ 3 $ & $ 63.1744 $ & $ 0.06 $ & $ 0.12 $ & $ -0.0049 $ & $
0.0033 $ & $ -0.0018 $ & $ 0.0036 $ \\
$ 5 $ & $ 68.1156 $ & $ -0.003 $ & $ 0.095 $ & $ -0.0015 $ & $
0.0026 $ & $ -0.0006 $ & $ 0.0028 $ \\
$ 6 $ & $ 70.5862 $ & $ 0.07 $ & $ 0.11 $ & $ -0.0028 $ & $
0.0034 $ & $ -0.0009 $ & $ 0.0035 $ \\
$ 7 $ & $ 73.0568 $ & $ 0.159 $ & $ 0.067 $ & $ 0.0185 $ & $
0.0021 $ & $ -0.0296 $ & $ 0.0021 $ \\
$ 8 $ & $ 75.5274 $ & $ 0.19 $ & $ 0.11 $ & $ 0.0039 $ & $ 0.0039
$ & $ -0.0016 $ & $ 0.0035 $ \\
$ 9 $ & $ 77.9981 $ & $ 0.06 $ & $ 0.11 $ & $ -0.0086 $ & $
0.0042 $ & $ 0.0064 $ & $ 0.0038 $ \\
$ 10 $ & $ 80.4687 $ & $ -0.074 $ & $ 0.072 $ & $ 0.0037 $ & $
0.0026 $ & $ -0.0108 $ & $ 0.0024 $ \\

\hline
\end{tabular}
\tablecomments{
$^a$Transit number. 
$^b$Mid transit time in BJD -- 2454900. 
$^c$O-C time difference. 
$^d$O-C uncertainty. 
$^e$Fractional duration variation: (duration of transit -- average)/average. 
$^f$TDV uncertainty. 
$^g$Fractional depth variation. 
$^h$TPV uncertainty.  
\\
(This table is available in its entirety in a machine-readable
form in ftp://wise-ftp.tau.ac.il/pub/tauttv/TTV. A portion of the
table of KOI-$1.01$ is
shown here for guidance regarding its form and content.)}

\label{tab:TTV}
\end{table}

\section{Identifying  KOIs with significant TTVs}
\label{interesting}

As the main focus of this study is the TTVs of the KOIs, the
next sections concentrate on the analysis of the derived O-Cs.
Analyses of duration (TDV) and depth (TPV) variations are deferred to a later paper.

In order to identify KOIs with significant TTVs, we computed a
few
statistics
(see F11, F12 and S12) to characterize the scatter of the
derived O-Cs. We obtained these statistics, listed in Table~\ref{tab:TTV_master}, only for
$1897$  KOIs which had at least seven timing
measurements.

\begin{itemize}

\item

For each KOI, we list the scatter of the O-Cs,  $s_{\rm O-C}$,
which we defined as the median absolute
deviation (MAD) of the O-C series, and
$\overline{\sigma}_{\rm TT}$, their median error (see
Figure~\ref{MADvsMed_error} and the discussion there). High
values of  $s_{\rm O-C}$ relative to
$\overline{\sigma}_{\rm TT}$
may indicate a significant TTV, especially because the MAD
statistic is less sensitive to outliers than the r.m.s. 
\end{itemize}

However, the derived
ratio relies on our estimate of the timing error, which by itself
depends on the estimated error and the nature of the noise of the
{\it Kepler} measurements.
Although Figure~\ref{MADvsMed_error} and
Figure~\ref{Med_errors_vsSNR} indicate that our error estimates are realistic, we are not sure how accurate the uncertainties
for a given KOI are, because
of the unknown contribution of the
red noise in the {\it Kepler} data. 
Another drawback of the scatter/error ratio is its
insensitivity to the
order of the residuals. That is, any permutation of the residuals
yields the same two values. However, as pointed out by F11 \citep[see
also][]{agol05,holman05,lithwick12}, the expected time scale of the dynamical
interaction between planets is in most cases larger than the
orbital period of the transiting planet. We therefore can assume
long-term correlation in the planet's O-Cs, if indeed the planet
is subject to a dynamical perturbation.

We therefore do not rely solely on the $s_{\rm O-C}/\overline{\sigma}_{\rm TT}$ ratio,
and add three statistics that can indicate long-term
correlation of the O-Cs:
\begin{itemize}
\item
The Lomb-Scargle (LS) periodogram (e.g., S12), which searched for
a cosine-shape periodicity in the series of O-Cs. We identified the
highest peak in the periodogram and assigned a false-alarm
probability (FAP) to
the existence of the associated periodicity in the data. This was
done by calculating similar $10^4$ LS periodograms with different
random permutations of the same O-Cs, and obtaining the highest peak in
each of these periodograms. 
Table~\ref{tab:TTV_master}
quotes the estimated period and its FAP p-value.
\item
A long-term polynomial fit to the series of TTVs. A good
polynomial fit
usually indicates a long-term modulation with a time scale longer
than the
data span. We searched for a polynomial with a degree lower than
four, chose the
best fit and tested its significance with the $\mathcal{F}$-test
(e.g., F11). 
Table~\ref{tab:TTV_master} quotes the best polynomial fit and its FAP p-value.

\item The 'alarm' score $\mathcal{A}$ of \citet{tamuz06}, which is
sensitive to the correlation between adjacent O-Cs. The value of
$\mathcal{A}$ reflects the number of consecutive TTVs with the
same sign, without assuming any functional shape of the
modulation
(see \citet{tamuz06} for a detailed discussion). We calculated 
$\mathcal{A}$ 
relative to the assumption of no TTV. We assigned a
false-alarm probability to the occurrence of the obtained score
 by calculating alarm scores for $10^4$ different random
permutations of the same TTVs. 
Table~\ref{tab:TTV_master} quotes the alarm score
and its p-value.

\end{itemize}

Table~\ref{tab:TTV_master} can be used to identify KOIs with
significant TTVs of various time scales.


\begin{table}[h]
\footnotesize \caption{Statistical parameters of the O-Cs series
of Kepler
KOIs}
\begin{tabular}{|r|rr|rcr|rr|rr|}
\hline \hline

KOI & $\overline{\sigma}_{\rm TT}\tablenotemark{a} $ & $S_{\rm O-C}\tablenotemark{b} $& LS~~ & LS& p-LS\tablenotemark{e}  &
$\mathcal{A}$\tablenotemark{f}~~  & p-$\mathcal{A}$\tablenotemark{g}  & Pol.  & p-$\mathcal{F}$\tablenotemark{i} \\
     &  &  &        Period\tablenotemark{c}   & Peak\tablenotemark{d}   &   &     &    &  Deg.\tablenotemark{h}  &
\\
     & [min] & [min] &       [d]  ~~   &    & [log]        &
& [log]  &
&  [log]   \\

\hline
$ 1.01 $ & $ 0.09 $ & $ 0.09 $ & $ 195.56 $ & $ 6.05 $ & $ -0.2 $ & $ 0.282 $ & $ -1.3 $ & $ 1 $ & $ -0.3 $ \\ 
$ 2.01 $ & $ 0.25 $ & $ 0.24 $ & $ 21.55 $ & $ 12.72 $ & $ -2.7 $ & $ 0.121 $ & $ -1.0 $ & $ 2 $ & $ -0.3 $ \\ 
$ 3.01 $ & $ 0.21 $ & $ 0.31 $ & $ 73.21 $ & $ 4.84 $ & $ -0.1 $ & $ 0.239 $ & $ -1.0 $ & $ 3 $ & $ -0.3 $ \\ 
$ 4.01 $ & $ 2.22 $ & $ 2.79 $ & $ 11.96 $ & $ 5.82 $ & $ -0.2 $ & $ -0.241 $ & $ -0.2 $ & $ 1 $ & $ -0.3 $ \\ 
$ 5.01 $ & $ 1.66 $ & $ 1.58 $ & $ 44.69 $ & $ 6.44 $ & $ -0.5 $ & $ 0.223 $ & $ -1.1 $ & $ 1 $ & $ -0.3 $ \\ 
$ 7.01 $ & $ 3.59 $ & $ 3.37 $ & $ 6.89 $ & $ 4.79 $ & $ 0.0 $ & $ -0.144 $ & $ -0.2 $ & $ 1 $ & $ -0.3 $ \\ 
$ 10.01 $ & $ 0.57 $ & $ 0.56 $ & $ 16.92 $ & $ 6.68 $ & $ -0.5 $ & $ -0.256 $ & $ -0.1 $ & $ 2 $ & $ -0.3 $ \\ 
$ 12.01 $ & $ 0.69 $ & $ 1.33 $ & $ 849.19 $ & $ 8.84 $ & $ -2.4 $ & $ 2.295 $ & $ -3.5 $ & $ 2 $ & $ -0.5 $ \\ 
$ 13.01 $ & $ 0.18 $ & $ 0.15 $ & $ 5.72 $ & $ 15.44 $ & $ -4.0 $ & $ -0.082 $ & $ -0.4 $ & $ 2 $ & $ -0.3 $ \\ 
$ 17.01 $ & $ 0.33 $ & $ 0.37 $ & $ 10.89 $ & $ 5.97 $ & $ -0.2 $ & $ 0.692 $ & $ -2.2 $ & $ 1 $ & $ -0.3 $ \\

\hline
\end{tabular}
\tablecomments{
$^a$O-C uncertainty median. 
$^b$O-C scatter (1.483 times the MAD).
$^c$Lomb-Scargle highest-peak period. 
$^d$The height of the Lomb-Scargle highest peak. 
$^e$The logarithmic of the p-value of the $\mathcal{F}$-test for the highest LS peak found. 
$^f$Alarm score (see text). 
$^g$The logarithmic of the p-value of the alarm found.  
$^h$Best fitted polynomial degree. 
$^i$The logarithmic of the p-value of the $\mathcal{F}$-test for the best polynomial fit. \\
(This table is available in its entirety in a machine-readable form
in ftp://wise-ftp.tau.ac.il/pub/tauttv/TTV. A portion is shown
here for guidance regarding its form and content.)}

\label{tab:TTV_master}
\end{table}

\section{KOIs with significant TTVs}
\label{long}

\begin{figure*}[p]
\centering
\resizebox{16cm}{11cm}
{\includegraphics{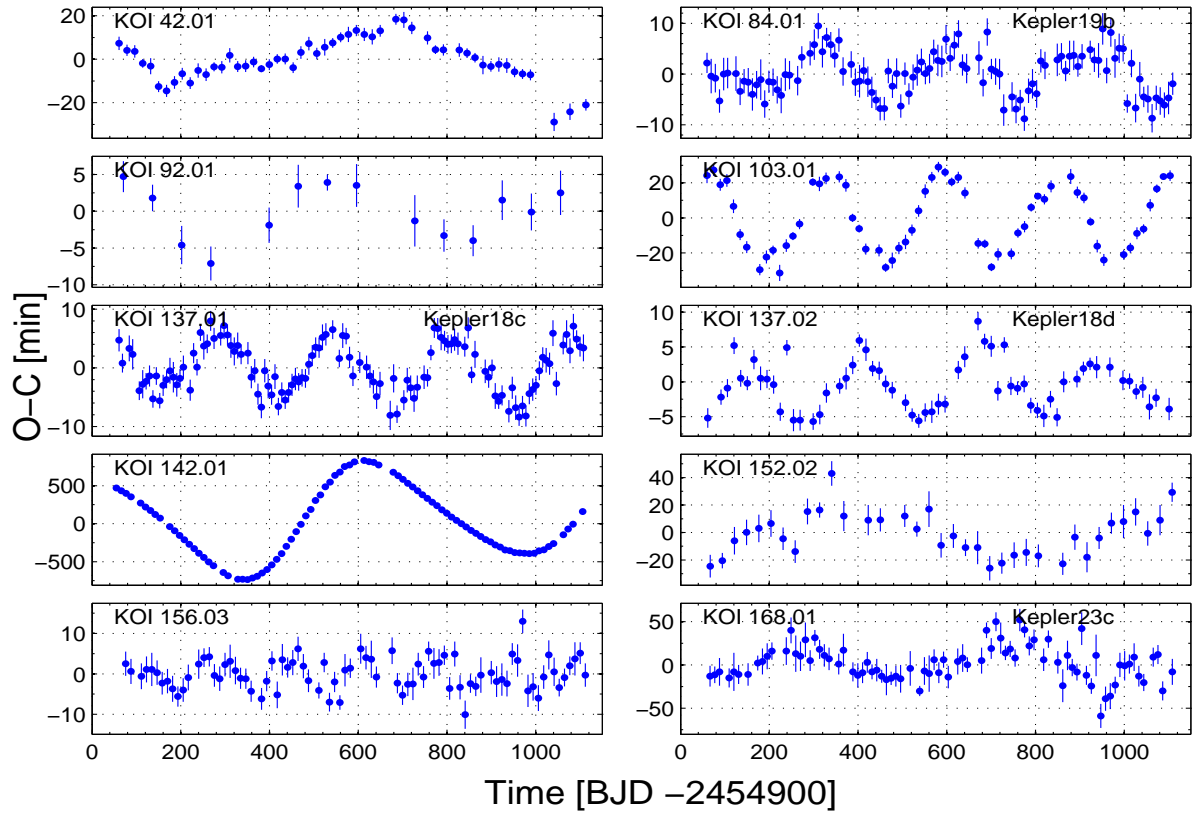}}
\caption{KOIs with significant TTVs.
}
\label{TTV1}
\end{figure*}

\begin{figure*}[p]
\centering
\resizebox{16cm}{11cm}
{\includegraphics{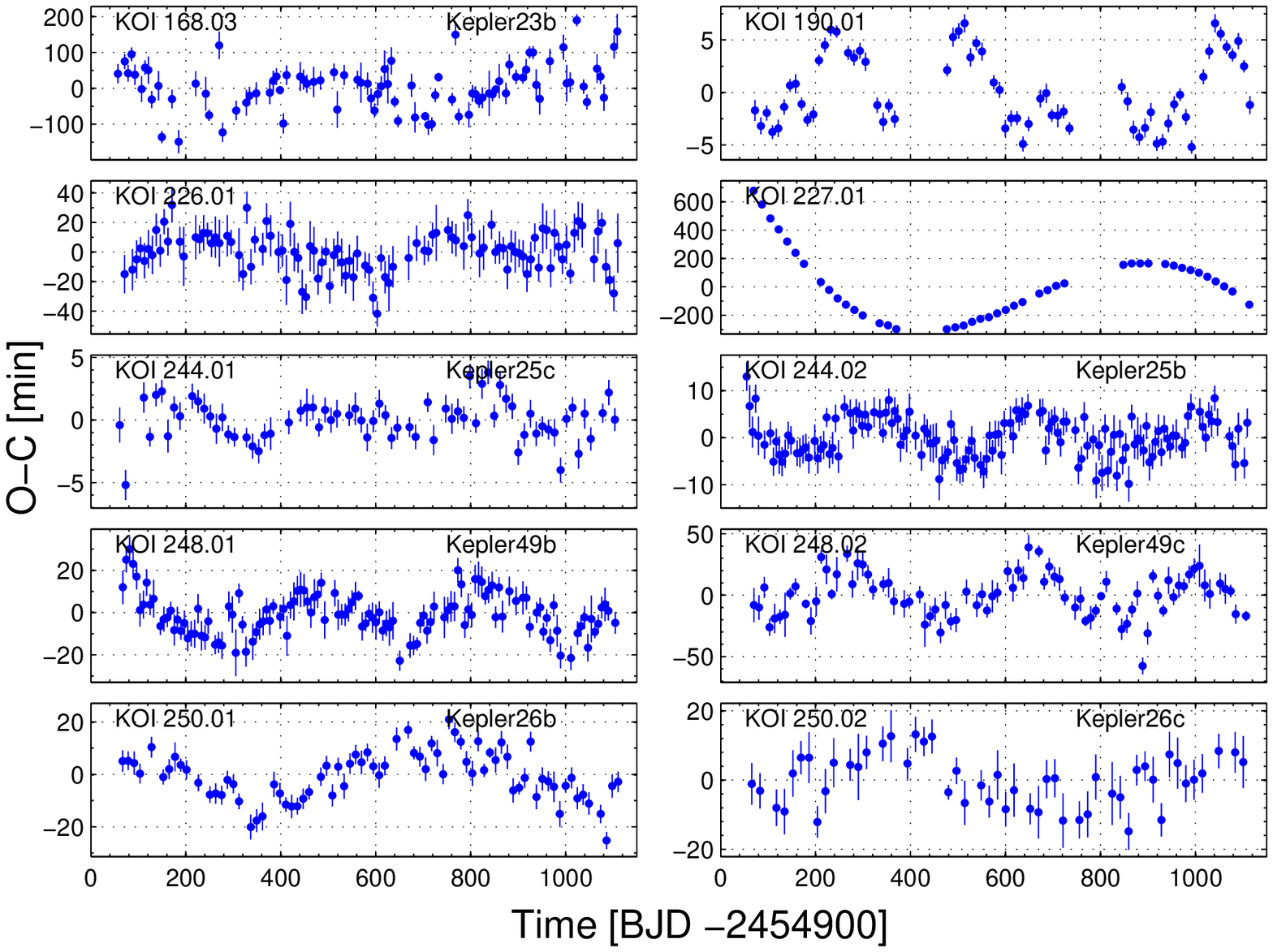}}
\caption{KOIs with significant TTVs.
}
\label{TTV2}
\end{figure*}

\begin{figure*}[p]
\centering
\resizebox{16cm}{11cm}
{\includegraphics{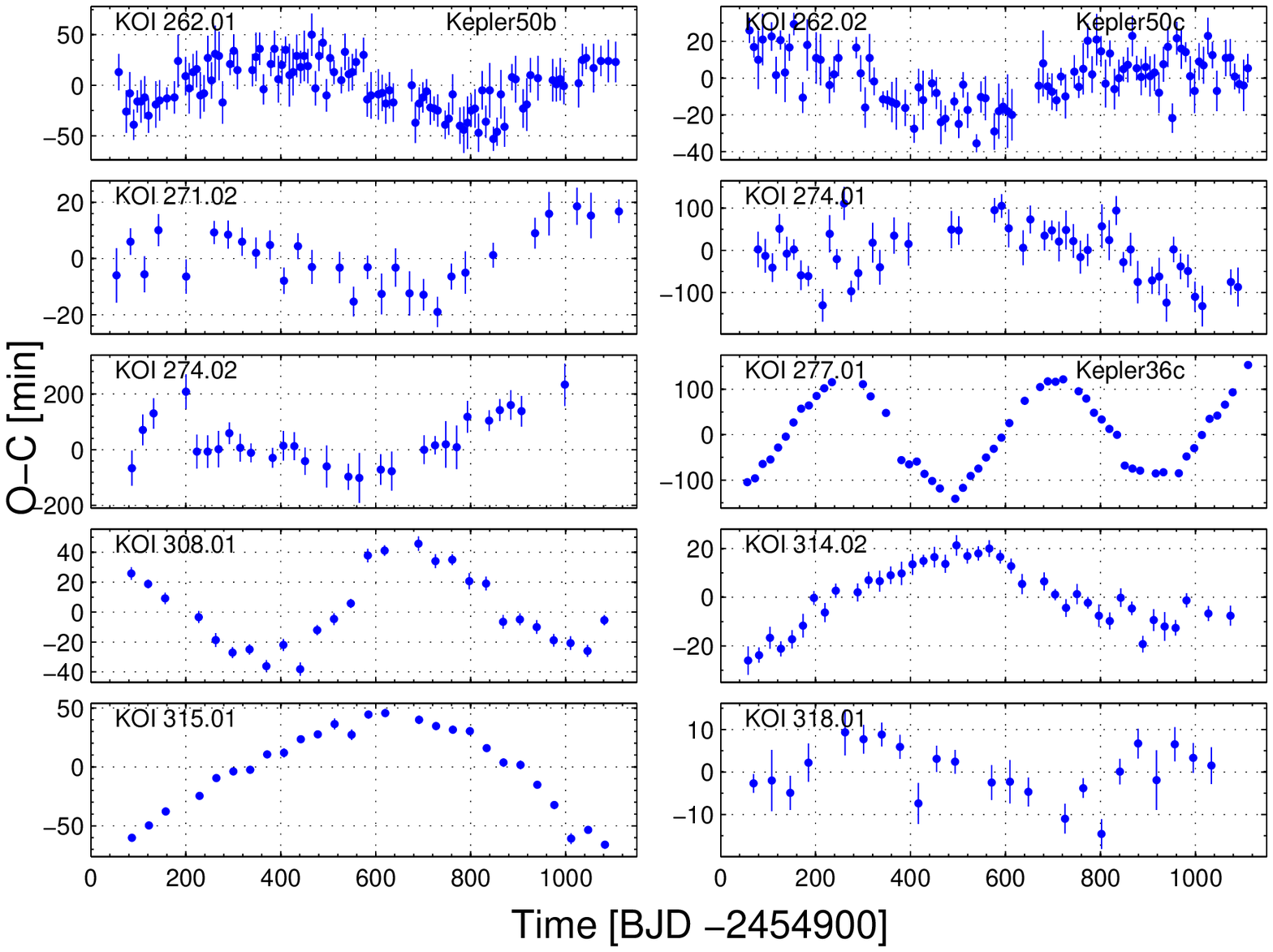}}
\caption{KOIs with significant TTVs.
}
\label{TTV3}
\end{figure*}

\begin{figure*}[p]
\centering
\resizebox{16cm}{11cm}
{\includegraphics{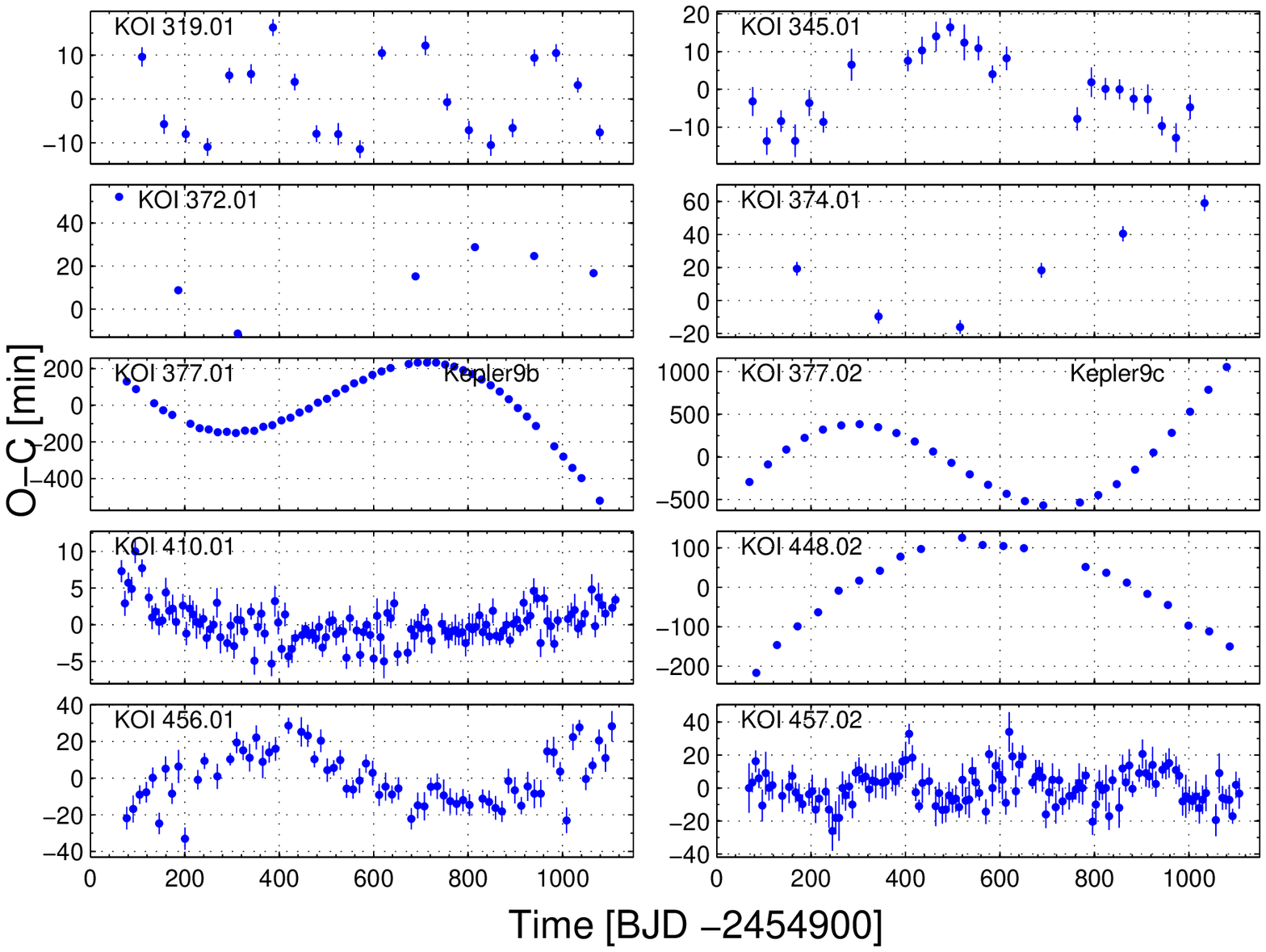}}
\caption{KOIs with significant TTVs.
}
\label{TTV4}
\end{figure*}

\begin{figure*}[p]
\centering
\resizebox{16cm}{11cm}
{\includegraphics{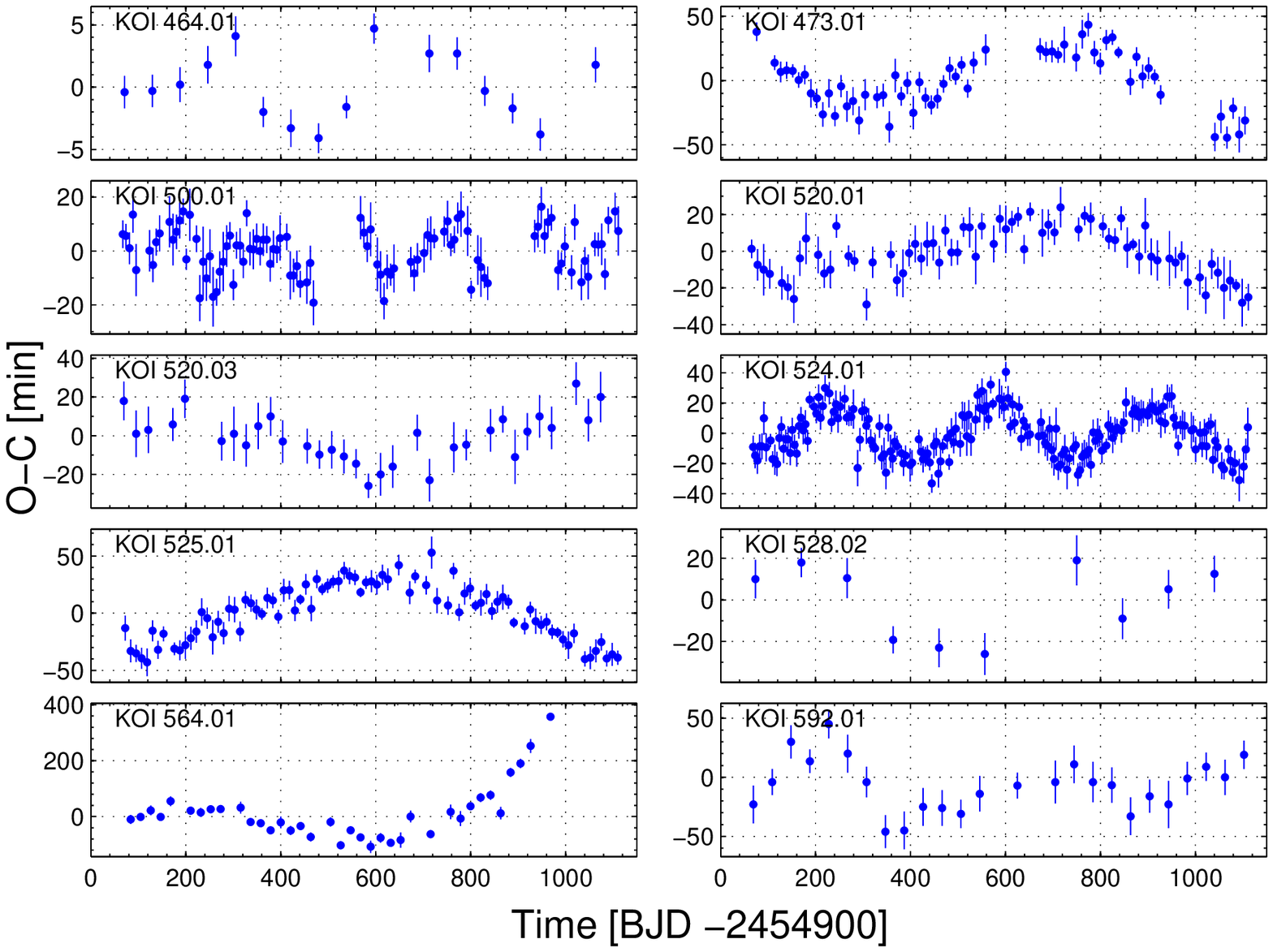}}
\caption{KOIs with significant TTVs.
}
\label{TTV5}
\end{figure*}

\begin{figure*}[p]
\centering
\resizebox{16cm}{11cm}
{\includegraphics{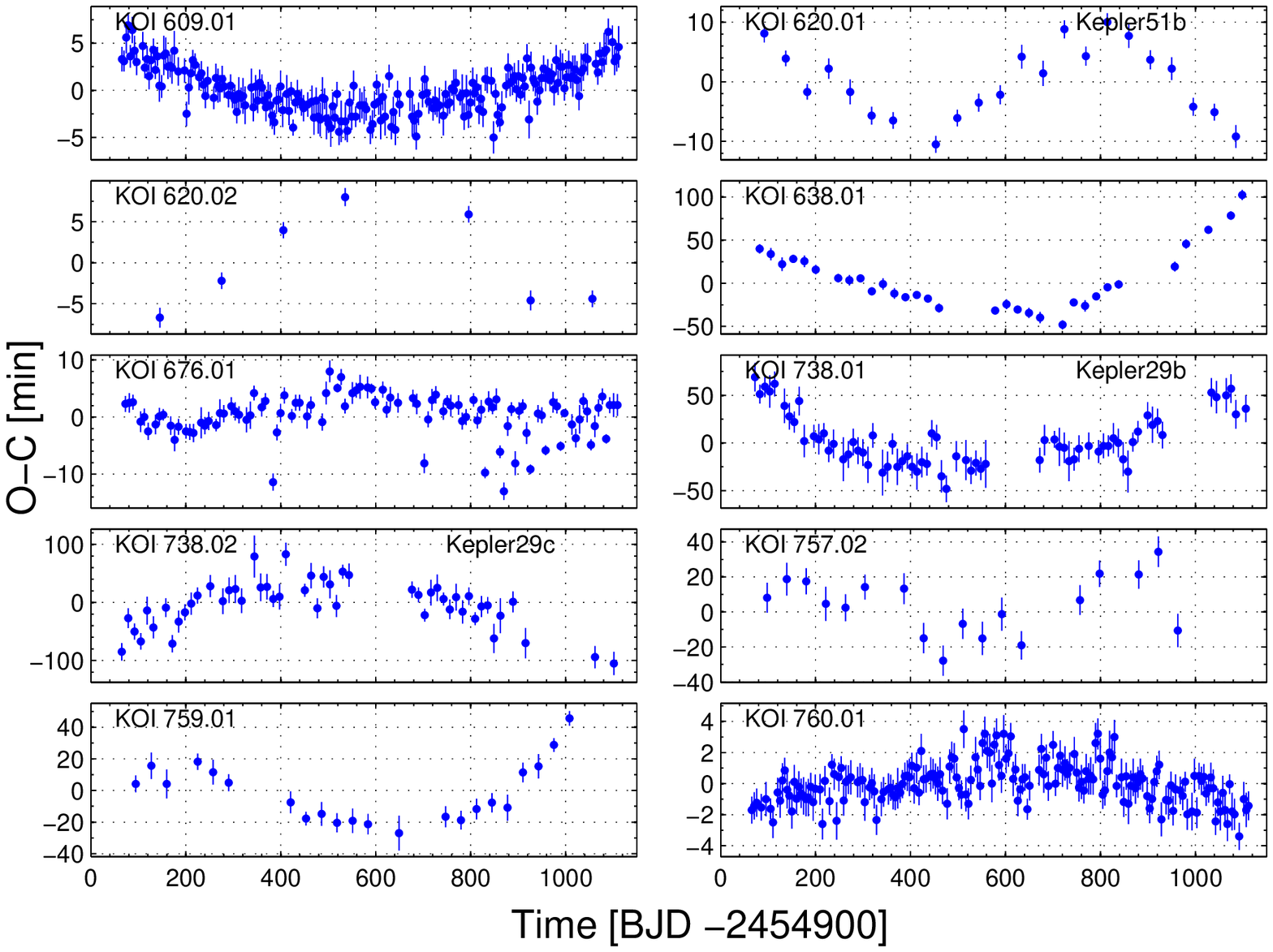}}
\caption{KOIs with significant TTVs.
}
\label{TTV6}
\end{figure*}

\begin{figure*}[p]
\centering
\resizebox{16cm}{11cm}
{\includegraphics{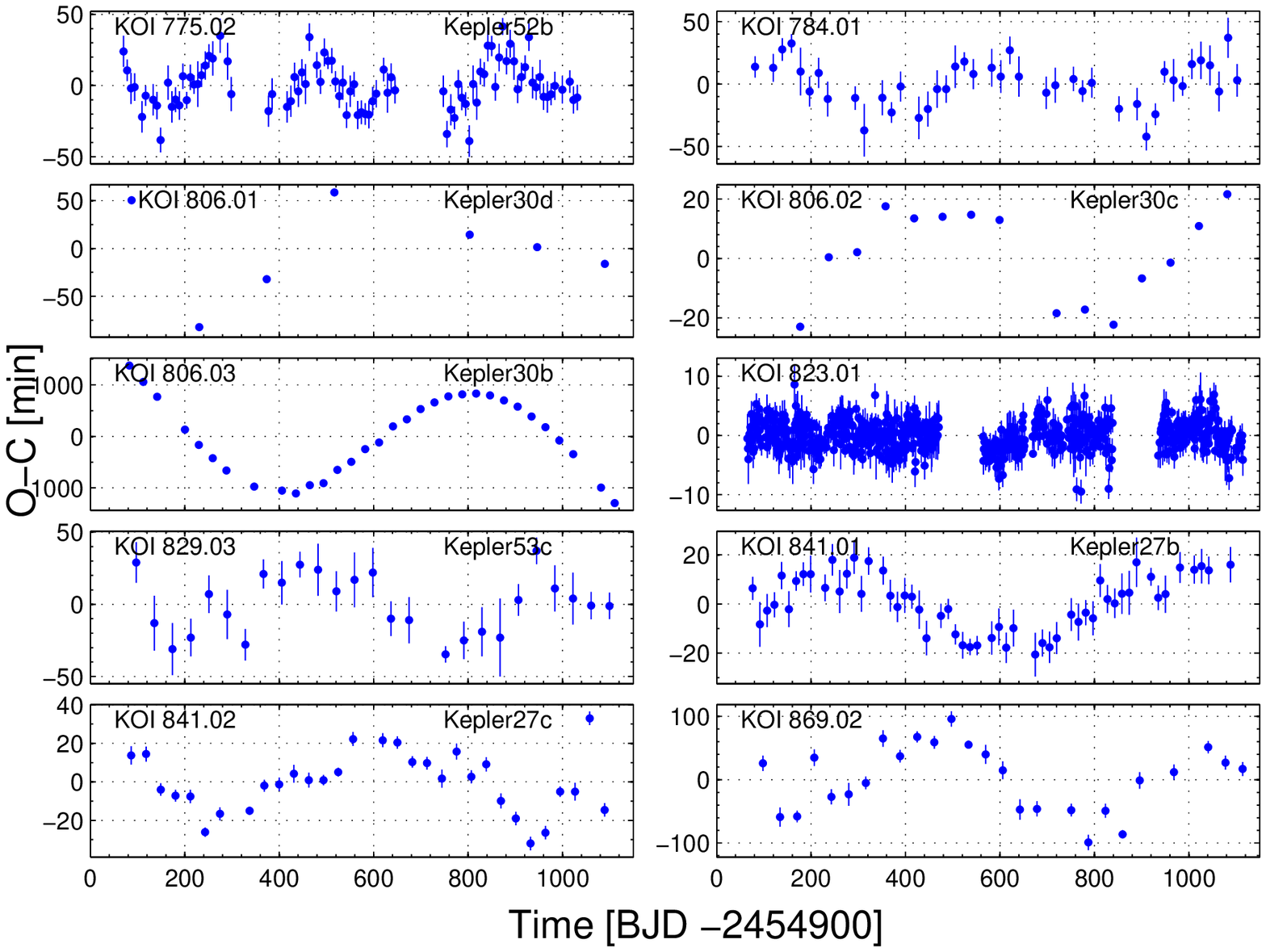}}
\caption{KOIs with significant TTVs.
}
\label{TTV7}
\end{figure*}

\begin{figure*}[p]
\centering
\resizebox{16cm}{11cm}
{\includegraphics{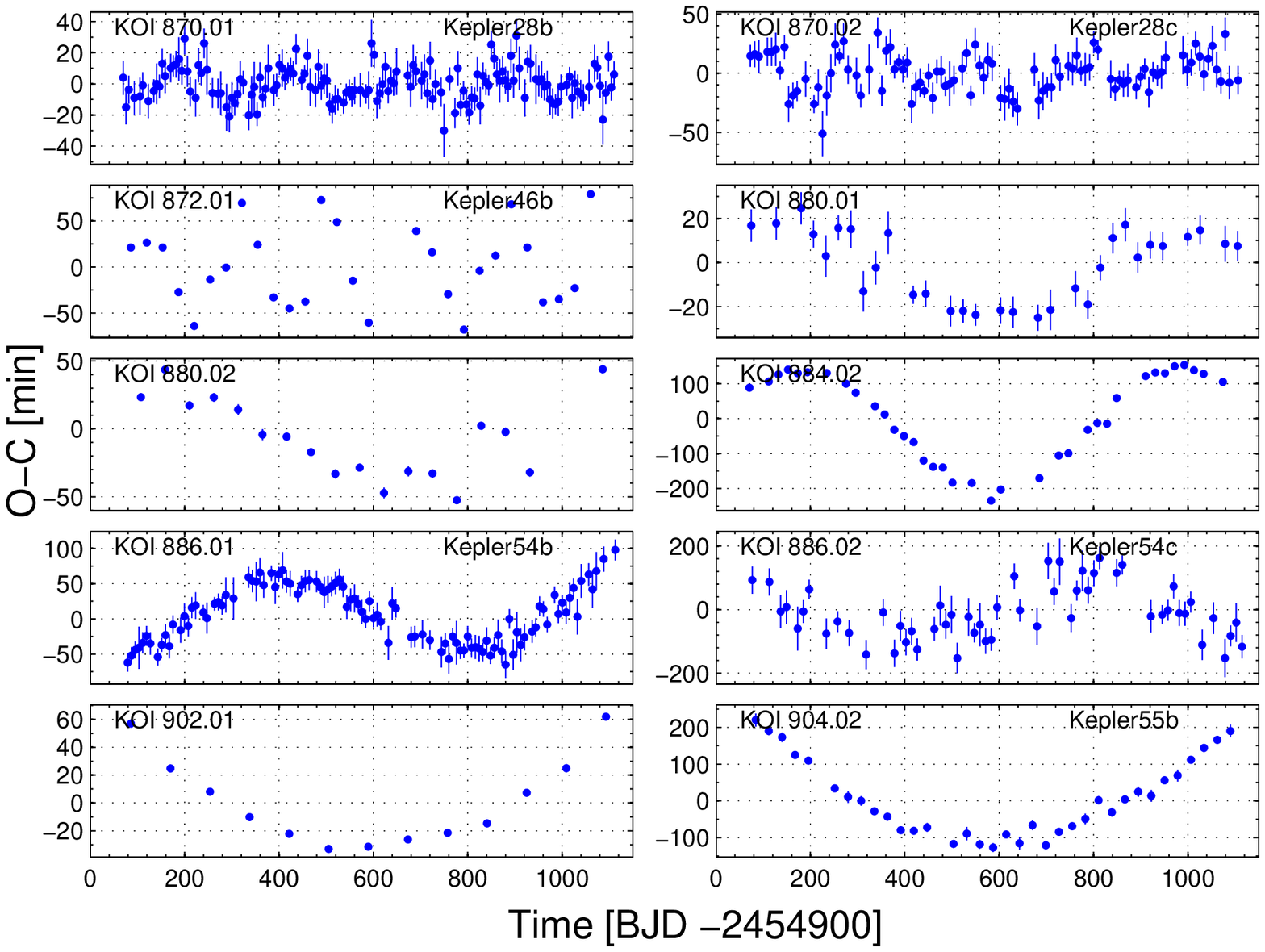}}
\caption{KOIs with significant TTVs.
}
\label{TTV8}
\end{figure*}

\begin{figure*}[p]
\centering
\resizebox{16cm}{11cm}
{\includegraphics{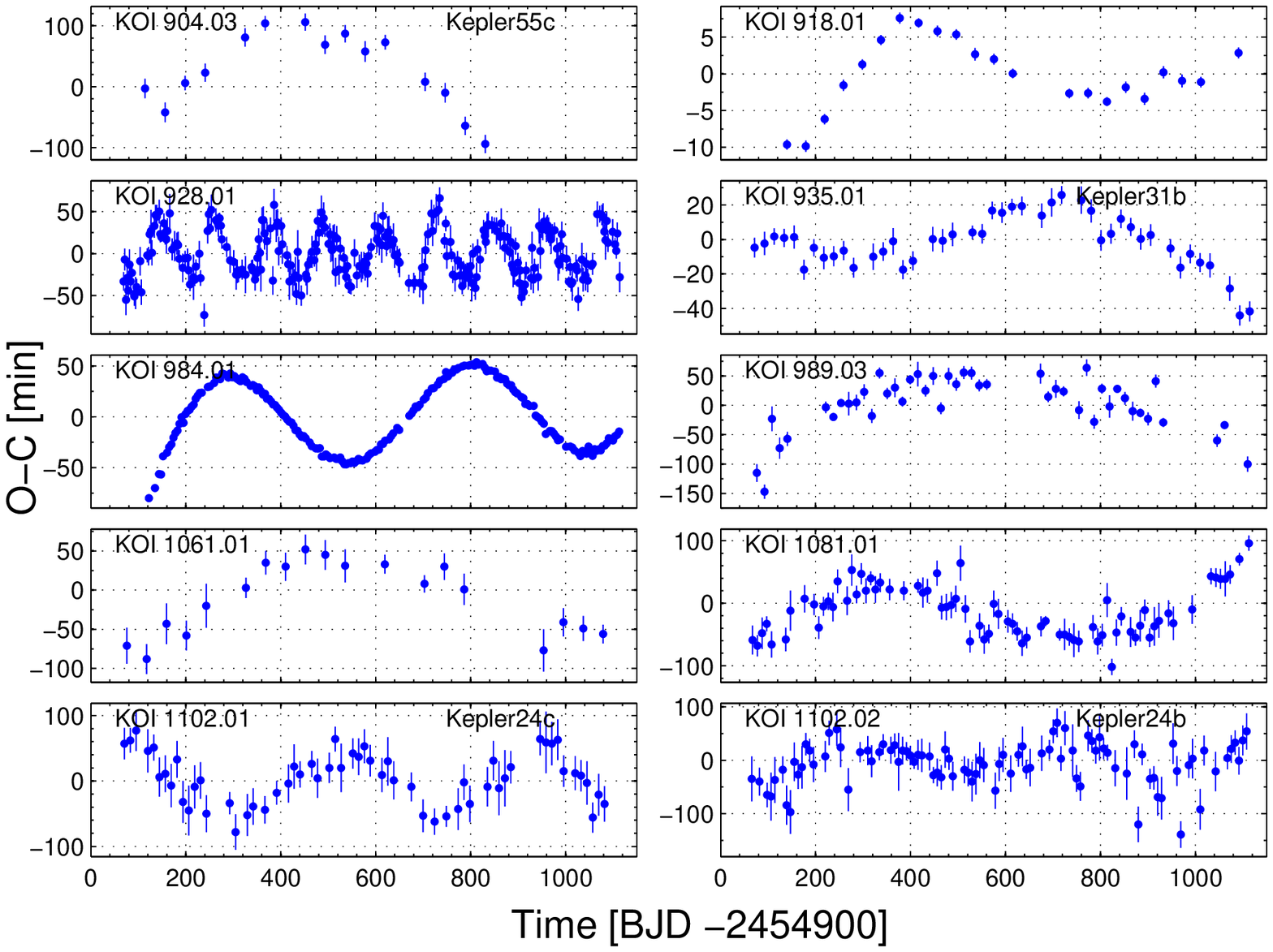}}
\caption{KOIs with significant TTVs.
}
\label{TTV9}
\end{figure*}

\begin{figure*}[p]
\centering
\resizebox{16cm}{11cm}
{\includegraphics{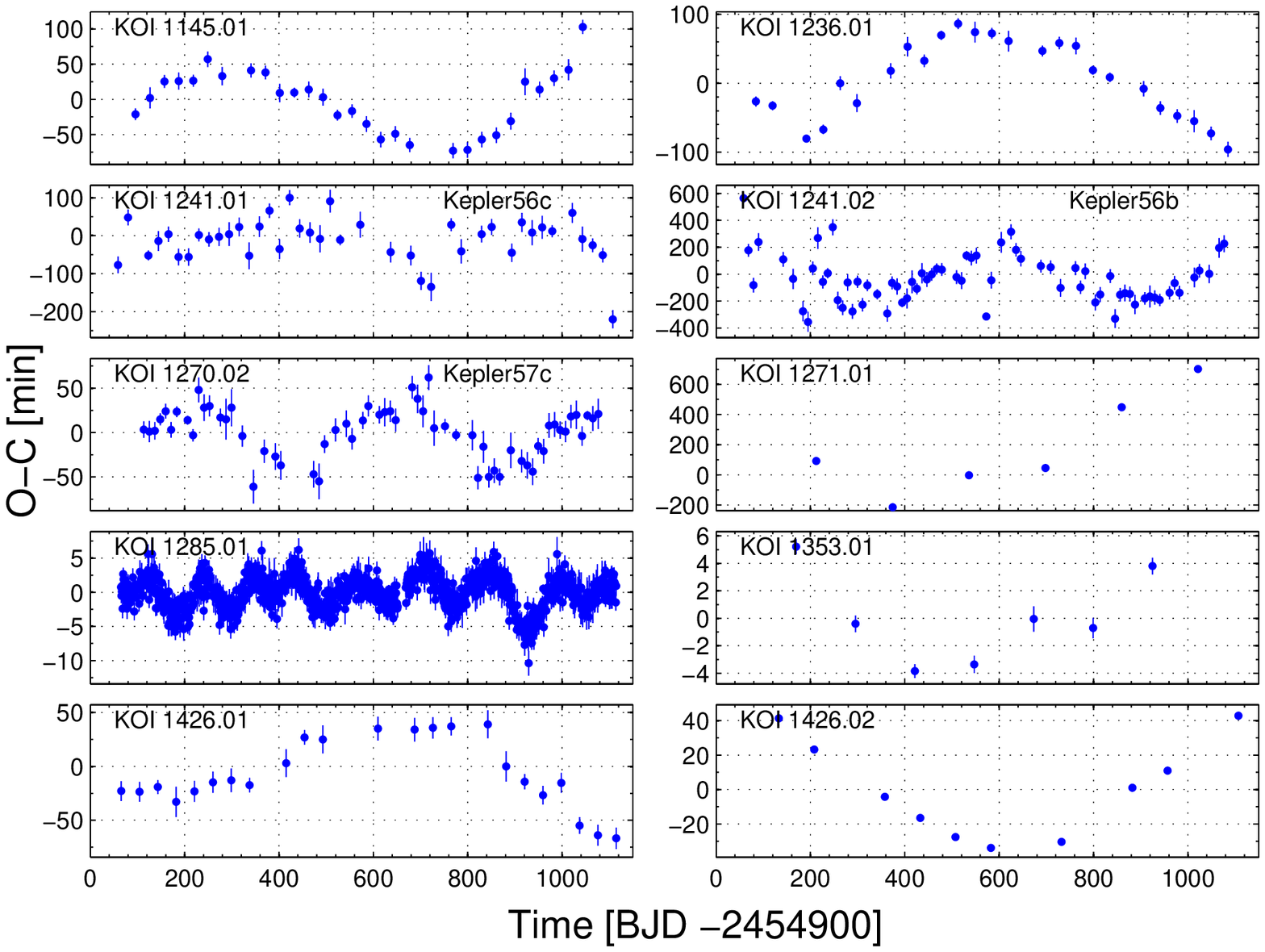}}
\caption{KOIs with significant TTVs.
}
\label{TTV10}
\end{figure*}

\begin{figure*}[p]
\centering
\resizebox{16cm}{11cm}
{\includegraphics{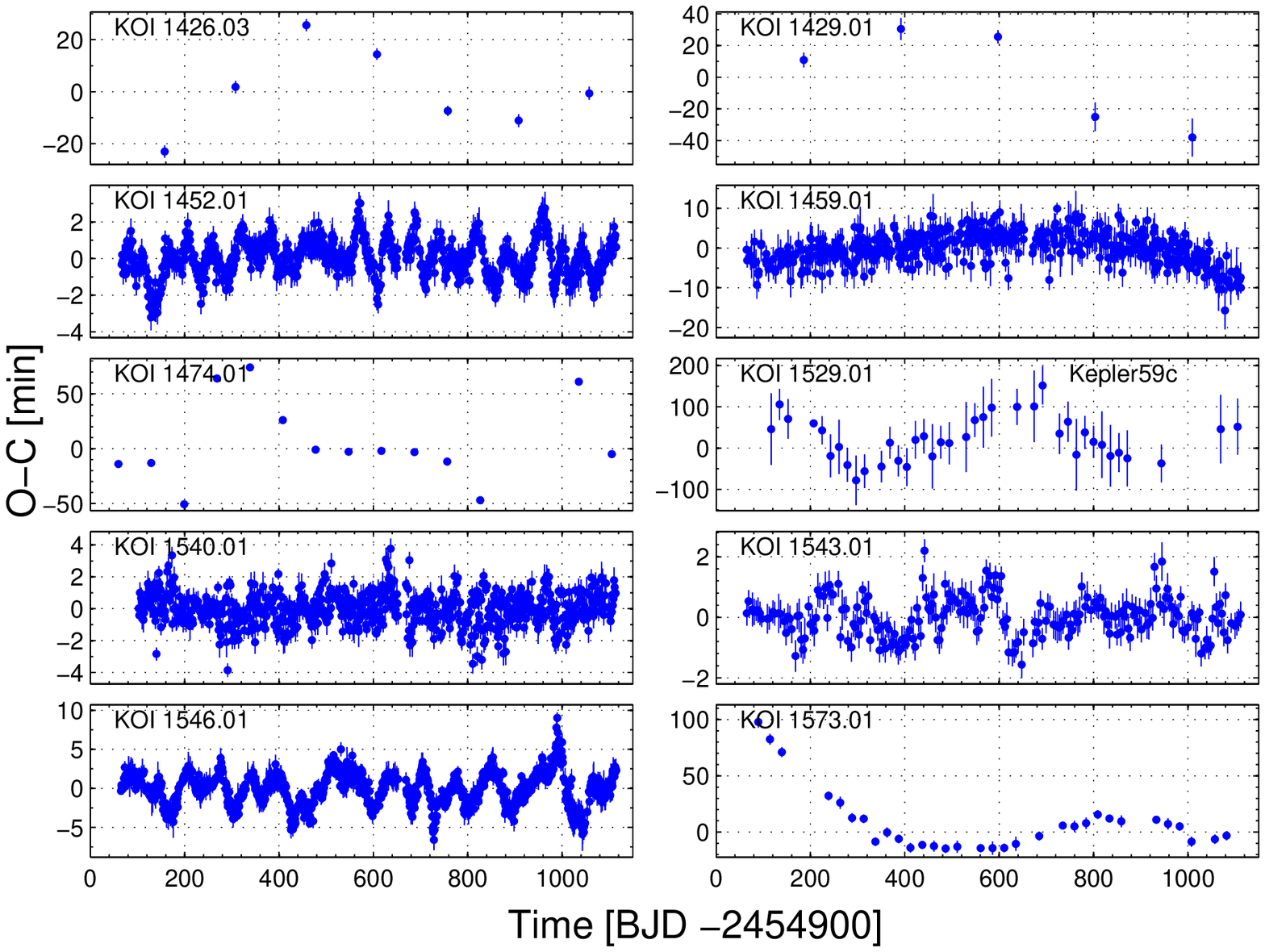}}
\caption{KOIs with significant TTVs.
}
\label{TTV11}
\end{figure*}

\begin{figure*}[p]
\centering
\resizebox{16cm}{11cm} {\includegraphics{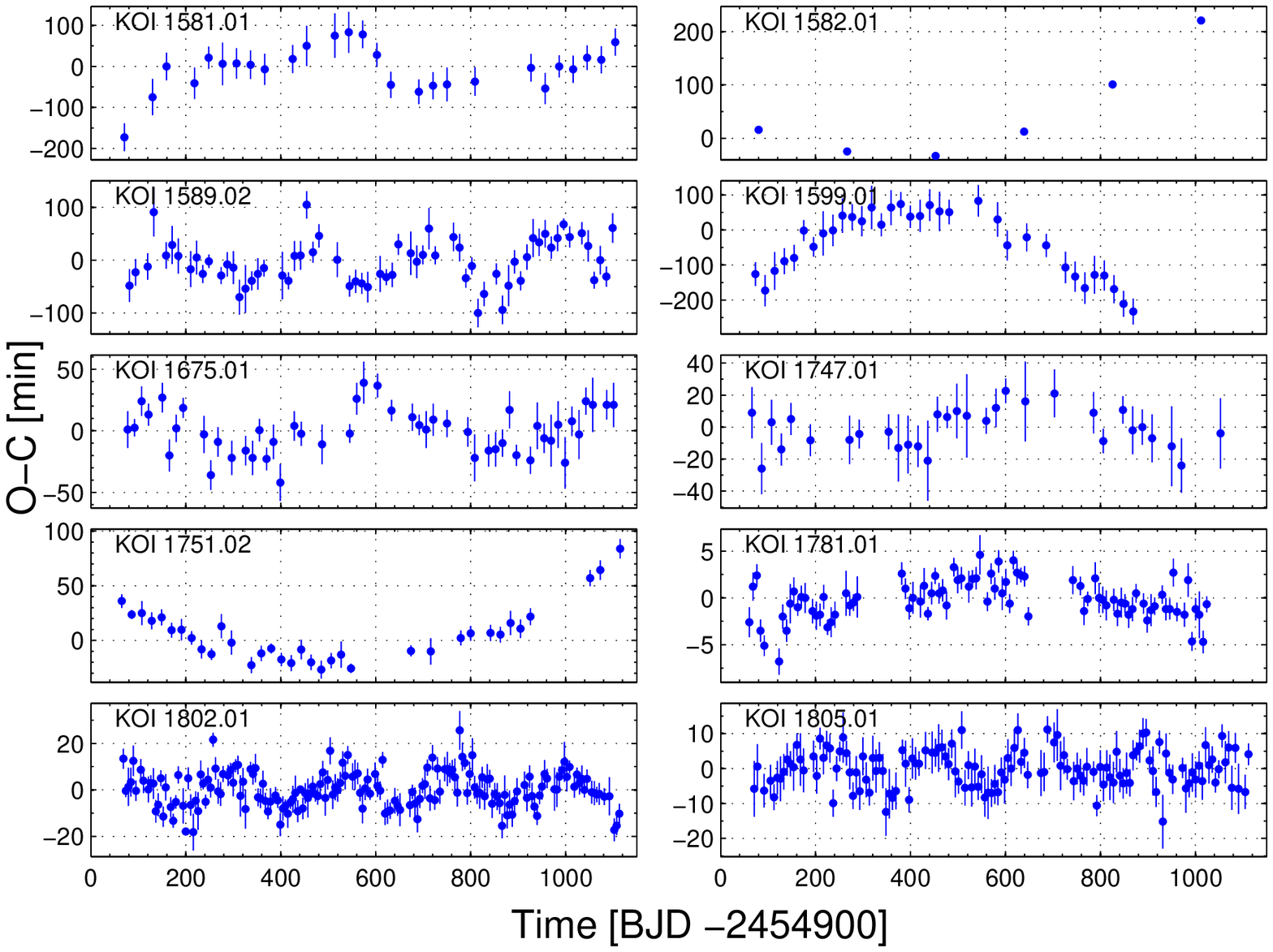}}
\caption{KOIs with significant TTVs.
}
\label{TTV12}
\end{figure*}

\begin{figure*}[p]
\centering
\resizebox{16cm}{11cm} {\includegraphics{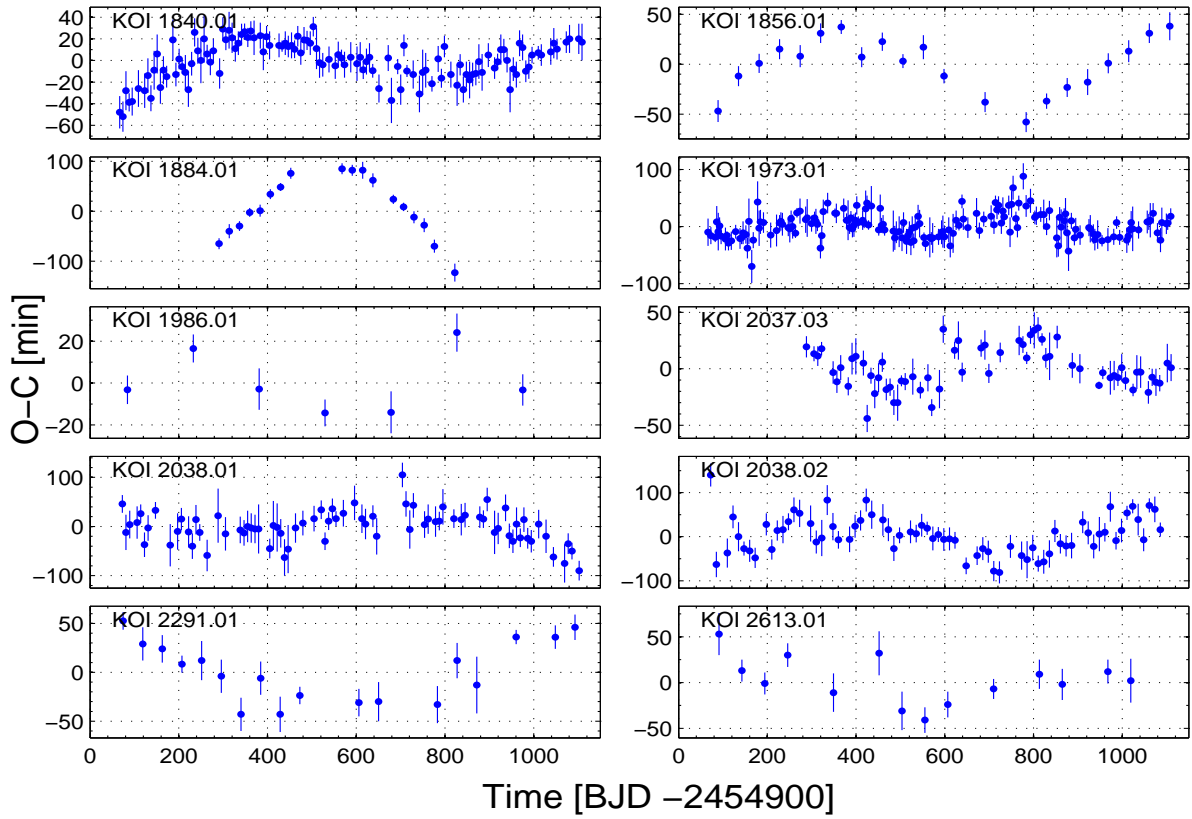}}
\caption{KOIs with significant TTVs.
}
\label{TTV13}
\end{figure*}

In this section we single out 130 systems with
significant TTVs, either because they have large scatter 
($s_{\rm O-C}$ /$\overline{\sigma}_{\rm TT} > 15$), display a periodic modulation (LS FAP lower than
$3\times$10$^{-4}$) or show a parabolic trend (see Table~\ref{tab:TTV_master}).
 Figures~\ref{TTV1}--\ref{TTV13} display the O-Cs of these systems,
and Table~\ref{tab:TTV_Interesting} summarizes 
their variability features.
Eight KOIs --- 94.02, 341.01, 1376.01, 1458.01, 1814.01, 1815.01, 2276.01 and 2631.01 are not included because they do not look significantly variable, even though they have passed one of these criteria. 

Eighty five of the $130$ KOIs showed
some periodicity, with time scales ranging from $100$ to
$1000$ days and amplitudes of $1$--$1000$ minutes.
For each of these $85$ systems, we derived a fit to the O-Cs (not
plotted but given in the table), composed of a straight line,
which could
present a correction to the orbital period of the transiting
planet, {\it together} with a cosine function with the
best-found period and phase.
Table~\ref{tab:TTV_Interesting} lists  the period and its error for $48$ KOIs. For $37$ systems the period
found
was too long or the fit was not good enough and we could not
derive its uncertainty. In those
cases, the period listed is just an approximation. In one special
case --- KOI-142.01, we fitted a straight line with {\it two}
different cosine functions.

For $39$ KOIs, the O-C series did not exhibit a maximum {\it and}
a minimum, and therefore we have not fitted a cosine function to the data. 
This probably meant that the time scale of the modulation was longer than
the time span of the data. In those cases, we fitted the O-Cs with
a long-term parabola only, and added a note in Table~\ref{tab:TTV_Interesting}.

For six systems ---
KOI-1285.01, 1452.01, 1474.01, 1540.01, 1543.01 and 1546.01, neither a cosine
function nor a parabola could be fitted, but the O-Cs looked
nevertheless significant (see Section~\ref{comments} for s short
discussion of all six KOIs).

Table~\ref{tab:TTV_Interesting} lists the KOI number, the
orbital period of the transiting planet and the model we used, either a
Cosine function, "C", or a polynomial "P". For a Cosine fit, we list the
TTV period and its error, when available, and the amplitude. The next column gives the scatter of the residuals relative
to the  found fit (which is not plotted). We also list the
number of TTV measurements, the multiplicity of the
of KOI and references to previous studies, when
available. In Section~\ref{comments} we briefly comment on some of the systems listed here. These systems are marked by an  asterisk in the table.


\begin{table}
\footnotesize
\caption{KOIs with significant TTV}
\begin{tabular}{|rr|lrrrrr|r|r|l|}
\hline \hline

KOI & Period\tablenotemark{a} & Model\tablenotemark{b}  & Period\tablenotemark{c}   & $\sigma_P\tablenotemark{d} $ & Amp\tablenotemark{e}   &$\sigma_A\tablenotemark{f} $&
Res\tablenotemark{g}  & $N$\tablenotemark{h}  &   Multi-  &  Ref.\tablenotemark{j} \\
     &  [d] &    & [d]      & [d]    &  [min]    &   [min]   &
[min] &  & plicity\tablenotemark{i}  & \\

\hline
$ 42.01 $ & $ 17.83 $ & C & $ 960 $ & \nodata & $ 13.91 $ & $
0.91 $ & $ 3.3 $ & $ 53 $ & $ 1 $ &  \\
$ 84.01 $ & $ 9.29 $ & C & $ 300 $ & $ 31 $ & $ 4.54 $ & $ 0.39 $
& $ 2.7 $ & $ 104 $ & $ 1 $ & \tablenotemark{3}Kepler19b \\
$ 92.01 $ & $ 65.70 $ & C & $ 519 $ & $ 84 $ & $ 4.42 $ & $ 0.76
$ & $ 2 $ & $ 14 $ & $ 1 $ &  \\
$ 103.01 $ & $ 14.91 $ & C & $ 261 $ & $ 13 $ & $ 26.14 $ & $
0.83 $ & $ 4.5 $ & $ 61 $ & $ 1 $ & \tablenotemark{1,2}~~~ \\
$ 137.01 $ & $ 7.64 $ & C & $ 268 $ & $ 21 $ & $ 5.38 $ & $ 0.26
$ & $ 1.7 $ & $ 120 $ & $ 3 $ & \tablenotemark{1,4}~~Kepler18c \\
$ 137.02 $ & $ 14.86 $ & C & $ 267 $ & $ 26 $ & $ 4.11 $ & $ 0.31
$ & $ 1.2 $ & $ 61 $ & $ 3 $ & \tablenotemark{1,2,4}~~~~Kepler18d
\\
$ \tablenotemark{*}142.01 $ & $ 10.95 $ & C & $ 618 $ & $ 58 $ &
$ 664 $ & $ 15 $ & $ 96 $ & $ 88 $ & $ 1 $ &
\tablenotemark{1,2,12}~~~ \\
                  &                 &     & $ 339 $ & $ 20 $ & $
111 $ & $ 5 $   &  $ 25 $ &           &          & \\
$ 152.02 $ & $ 27.40 $ & C & $ 870 $ & \nodata & $ 20.8 $ & $ 2.4
$ & $ 7.7 $ & $ 36 $ & $ 3 $ &  \tablenotemark{13,17} \\
$ 156.03 $ & $ 11.78 $ & C & $ 167 $ & $ 12 $ & $ 3.02 $ & $ 0.48
$ & $ 2.5 $ & $ 81 $ & $ 3 $ &  \\
$ 168.01 $ & $ 10.74 $ & C & $ 474 $ & $ 89 $ & $ 19.8 $ & $ 2.2
$ & $ 13 $ & $ 82 $ & $ 3 $ & \tablenotemark{2,5}~~Kepler23c \\
 \hline
$ 168.03 $ & $ 7.11 $ & C & $ 478 $ & $ 90 $ & $ 52 $ & $ 8.2 $ &
$ 38 $ & $ 88 $ & $ 3 $ & \tablenotemark{5}~~~Kepler23b \\
$  \tablenotemark{*}190.01 $ & $ 12.26 $ & C & $ 267 $ & $ 22 $ &
$ 4.31 $ & $ 0.31
$ & $ 1.6 $ & $ 65 $ & $ 1 $ &   \tablenotemark{14} \\
$ 226.01 $ & $ 8.31 $ & C & $ 610 $ & $ 210 $ & $ 8.9 $ & $ 1.8 $ & $ 12 $ & $ 105 $ & $ 1 $ &  \\ 
$ 227.01 $ & $ 17.70 $ & C & $ 1000 $ & \nodata & $ 397.4 $ & $
9.5 $ & $ 29 $ & $ 44 $ & $ 1 $ & \tablenotemark{1,2}~~~ \\
$ 244.01 $ & $ 12.72 $ & C & $ 316 $ & $ 39 $ & $ 1.24 $ & $ 0.22
$ & $ 1.2 $ & $ 71 $ & $ 2 $ & \tablenotemark{1,6}~~Kepler25c \\
$ 244.02 $ & $ 6.24 $ & C & $ 340 $ & $ 45 $ & $ 4.13 $ & $ 0.39
$ & $ 3 $ & $ 145 $ & $ 2 $ & \tablenotemark{1,2,6}~~~~Kepler25b
\\
$ 248.01 $ & $ 7.20 $ & C & $ 384 $ & $ 58 $ & $ 9.51 $ & $ 0.92
$ & $ 6.2 $ & $ 125 $ & $ 4 $ &
\tablenotemark{1,2,7,17}~~~~~~~Kepler49b \\

$ 248.02 $ & $ 10.91 $ & C & $ 370 $ & $ 54 $ & $ 17.7 $ & $ 1.8
$ & $ 11 $ & $ 83 $ & $ 4 $ &
\tablenotemark{1,7,17}~~~~~Kepler49c \\
$ 250.01 $ & $ 12.28 $ & C & $ 750 $ & \nodata & $ 10.35 $ & $
0.98 $ & $ 5.9 $ & $ 76 $ & $ 4 $ & \tablenotemark{6}Kepler26b \\
$ 250.02 $ & $ 17.25 $ & C & $ 800 $ & \nodata & $ 7.7 $ & $ 1.3
$ & $ 5.9 $ & $ 50 $ & $ 4 $ & \tablenotemark{1,6}~~Kepler26c \\

\hline
$ 262.01 $ & $ 7.81 $ & C & $ 750 $ & \nodata & $ 26.4 $ & $ 2.2
$ & $ 15 $ & $ 103 $ & $ 2 $ & \tablenotemark{7}Kepler50b \\
$ 262.02 $ & $ 9.38 $ & C & $ 880 $ & \nodata & $ 15.5 $ & $ 1.6
$ & $ 9.9 $ & $ 92 $ & $ 2 $ & \tablenotemark{7}Kepler50c \\
$ 271.02 $ & $ 29.39 $ & C & $ 880 $ & \nodata & $ 12.1 $ & $ 1.6
$ & $ 4.2 $ & $ 29 $ & $ 2 $ &  \tablenotemark{18}  \\
$ 274.01 $ & $ 15.09 $ & P & \nodata & \nodata & \nodata &
\nodata & $ 45 $ & $ 48 $ & $ 2 $ &  \\
$ 274.02 $ & $ 22.80 $ & P & \nodata & \nodata & \nodata &
\nodata & $ 45 $ & $ 29 $ & $ 2 $ &  \\
$ 277.01 $ & $ 16.23 $ & C & $ 440 $ & $ 29 $ & $ 116.6 $ & $ 3.1
$ & $ 15 $ & $ 54 $ & $ 1 $ &
\tablenotemark{1,2,8,18}~~~~~~~Kepler36c \\
$ 308.01 $ & $ 35.60 $ & C & $ 623 $ & $ 79 $ & $ 34.4 $ & $ 2.1
$ & $ 5.7 $ & $ 27 $ & $ 1 $ & \tablenotemark{2}~~~ \\
$ 314.02 $ & $ 23.09 $ & P & \nodata & \nodata & \nodata &
\nodata & $ 5.3 $ & $ 41 $ & $ 3 $ &  \\
$ 315.01 $ & $ 35.59 $ & P & \nodata & \nodata & \nodata &
\nodata & $ 8.5 $ & $ 27 $ & $ 1 $ &  \\
$ 318.01 $ & $ 38.58 $ & C & $ 690 $ & \nodata & $ 6.4 $ & $ 1.3
$ & $ 2.3 $ & $ 23 $ & $ 1 $ &  \\

\hline
\end{tabular}
\end{table}

\addtocounter{table}{-1}

\begin{table} \footnotesize \caption{ - continued}
\begin{tabular}{|rr|lrrrrr|r|r|l|}
\hline \hline

KOI & Period\tablenotemark{a} & Model\tablenotemark{b}  & Period\tablenotemark{c}   & $\sigma_P\tablenotemark{d} $ & Amp\tablenotemark{e}   &$\sigma_A\tablenotemark{f} $&
Res\tablenotemark{g}  & $N$\tablenotemark{h}  &   Multi-  &  Ref.\tablenotemark{j} \\
     &  [d] &    & [d]      & [d]    &  [min]    &   [min]   &
[min] &  & plicity\tablenotemark{i}  & \\

 \hline
$ 319.01 $ & $ 46.15 $ & C & $ 303 $ & $ 19 $ & $ 12.3 $ & $ 1.2
$ & $ 3.2 $ & $ 21 $ & $ 1 $ &  \\
$ 345.01 $ & $ 29.88 $ & P & \nodata & \nodata & \nodata &
\nodata & $ 3 $ & $ 24 $ & $ 1 $ &  \\
$ 372.01 $ & $ 125.63 $ & C & $ 1000 $ & \nodata & $ 35.8 $ & $ 5.6 $ & $ 7.6 $ & $ 7 $ & $ 1 $ &  \\ 

$ 374.01 $ & $ 172.69 $ & P & \nodata & \nodata & \nodata &
\nodata & $ 13 $ & $ 6 $ & $ 1 $ &  \\
$ 377.01 $ & $ 19.26 $ & C & $ 1000 $ & \nodata & $ 324.8 $ & $ 2
$ & $ 6.4 $ & $ 48 $ & $ 3 $ & \tablenotemark{2,9}~~Kepler9b \\
$ 377.02 $ & $ 38.88 $ & C & $ 1000 $ & \nodata & $ 764.9 $ & $
4.5 $ & $ 6.3 $ & $ 26 $ & $ 3 $ &
\tablenotemark{1,2,9}~~~~Kepler9c \\
$ 410.01 $ & $ 7.22 $ & P & \nodata & \nodata & \nodata & \nodata
& $ 1.7 $ & $ 130 $ & $ 1 $ &   \tablenotemark{14} \\
$ 448.02 $ & $ 43.59 $ & P & \nodata & \nodata & \nodata &
\nodata & $ 11 $ & $ 21 $ & $ 2 $ & \tablenotemark{2}~~~ \\
$ 456.01 $ & $ 13.70 $ & C & $ 730 $ & \nodata & $ 16.6 $ & $ 1.5
$ & $ 9.3 $ & $ 69 $ & $ 2 $ & \tablenotemark{2}~~~ \\
$ 457.02 $ & $ 7.06 $ & C & $ 281 $ & $ 36 $ & $ 8.5 $ & $ 1.1 $
& $ 7.4 $ & $ 128 $ & $ 2 $ &  \\
 \hline
$ 464.01 $ & $ 58.36 $ & C & $ 451 $ & $ 77 $ & $ 3.16 $ & $ 0.7
$ & $ 1.4 $ & $ 16 $ & $ 2 $ &  \\
$ 473.01 $ & $ 12.71 $ & C & $ 860 $ & \nodata & $ 30.4 $ & $ 2.1
$ & $ 12 $ & $ 60 $ & $ 1 $ & \tablenotemark{2}~~~ \\
$ 500.01 $ & $ 7.05 $ & C & $ 190 $ & $ 17 $ & $ 7.88 $ & $ 0.98
$ & $ 6.6 $ & $ 100 $ & $ 5 $ & \tablenotemark{1,17,19}~~~ \\
$ 520.01 $ & $ 12.76 $ & P & \nodata & \nodata & \nodata &
\nodata & $ 8.2 $ & $ 70 $ & $ 3 $ &  \\
$ 520.03 $ & $ 25.75 $ & P & \nodata & \nodata & \nodata &
\nodata & $ 7.1 $ & $ 32 $ & $ 3 $ &  \\
$ 524.01 $ & $ 4.59 $ & C & $ 336 $ & $ 44 $ & $ 16.98 $ & $ 0.82
$ & $ 7.9 $ & $ 195 $ & $ 1 $ & \tablenotemark{2}~~~ \\
$ 525.01 $ & $ 11.53 $ & P & \nodata & \nodata & \nodata &
\nodata & $ 9.8 $ & $ 84 $ & $ 1 $ &  \\
$ 528.02 $ & $ 96.68 $ & P & \nodata & \nodata & \nodata &
\nodata & $ 8 $ & $ 10 $ & $ 3 $ &  \\
$ 564.01 $ & $ 21.06 $ & C & $ 880 $ & \nodata & $ 120 $ & $ 11 $
& $ 24 $ & $ 37 $ & $ 2 $ & \tablenotemark{1,18}~~~ \\
$ 592.01 $ & $ 39.75 $ & C & $ 449 $ & $ 73 $ & $ 24.4 $ & $ 4.6
$ & $ 12 $ & $ 25 $ & $ 1 $ &  \\
\hline
$ \tablenotemark{*} 609.01 $ & $ 4.40 $ & P & \nodata & \nodata &
\nodata & \nodata
& $ 1.4 $ & $ 207 $ & $ 1 $ &    \tablenotemark{14} \\
$ 620.01 $ & $ 45.16 $ & C & $ 760 $ & \nodata & $ 8.5 $ & $ 0.78
$ & $ 2.3 $ & $ 22 $ & $ 3 $ & \tablenotemark{7}Kepler51b \\
$ 620.02 $ & $ 130.18 $ & P & \nodata & \nodata & \nodata &
\nodata & $ 1.2 $ & $ 7 $ & $ 3 $ &  \\
$ 638.01 $ & $ 23.64 $ & P & \nodata & \nodata & \nodata &
\nodata & $ 8.7 $ & $ 32 $ & $ 2 $ &  \\
$ 676.01 $ & $ 7.97 $ & C & $ 690 $ & \nodata & $ 2.48 $ & $ 0.36
$ & $ 2 $ & $ 102 $ & $ 2 $ &  \\
$ 738.01 $ & $ 10.34 $ & P & \nodata & \nodata & \nodata &
\nodata & $ 12 $ & $ 70 $ & $ 2 $ &
\tablenotemark{2,10}~~~Kepler29b \\
$ 738.02 $ & $ 13.29 $ & P & \nodata & \nodata & \nodata &
\nodata & $ 19 $ & $ 50 $ & $ 2 $ & \tablenotemark{10}~Kepler29c
\\
$ 757.02 $ & $ 41.19 $ & P & \nodata & \nodata & \nodata &
\nodata & $ 9.2 $ & $ 18 $ & $ 3 $ &  \\
$ 759.01 $ & $ 32.63 $ & P & \nodata & \nodata & \nodata &
\nodata & $ 11 $ & $ 22 $ & $ 1 $ &  \\
$ 760.01 $ & $ 4.96 $ & P & \nodata & \nodata & \nodata & \nodata
& $ 0.97 $ & $ 186 $ & $ 1 $ &  \\

\hline
\end{tabular}
\end{table}

\addtocounter{table}{-1}

\begin{table}\footnotesize \caption{ - continued}
\begin{tabular}{|rr|lrrrrr|r|r|l|}
\hline \hline

KOI & Period\tablenotemark{a} & Model\tablenotemark{b}  & Period\tablenotemark{c}   & $\sigma_P\tablenotemark{d} $ & Amp\tablenotemark{e}   &$\sigma_A\tablenotemark{f} $&
Res\tablenotemark{g}  & $N$\tablenotemark{h}  &   Multi-  &  Ref.\tablenotemark{j} \\
     &  [d] &    & [d]      & [d]    &  [min]    &   [min]   &
[min] &  & plicity\tablenotemark{i}  & \\

 \hline
$ 775.02 $ & $ 7.88 $ & C & $ 209 $ & $ 16 $ & $ 16.5 $ & $ 1.8 $
& $ 11 $ & $ 90 $ & $ 3 $ & \tablenotemark{7}Kepler52b \\
$ 784.01 $ & $ 19.27 $ & C & $ 486 $ & $ 88 $ & $ 18.4 $ & $ 2.8
$ & $ 14 $ & $ 42 $ & $ 2 $ & \tablenotemark{2}~~~ \\
$ 806.01 $ & $ 143.21 $ & C & $ 400 $ & \nodata & $ 60 $ & $ 20 $
& $ 30 $ & $ 7 $ & $ 3 $ & \tablenotemark{2,10,15}~~~~~~Kepler30d
\\
$ 806.02 $ & $ 60.32 $ & C & $ 750 $ & \nodata & $ 22.1 $ & $ 2.2
$ & $ 5.9 $ & $ 15 $ & $ 3 $ & \tablenotemark{10,15}~~~~Kepler30c
\\
$ 806.03 $ & $ 29.37 $ & C & $ 930 $ & $ 120 $ & $ 1343.5 $ & $
8.3 $ & $ 22 $ & $ 32 $ & $ 3 $ &
\tablenotemark{2,10,15}~~~~~~Kepler30b \\
$ \tablenotemark{*} 823.01 $ & $ 1.03 $ & C & $ 184 $ & $ 12 $ &
$ 0.85 $ & $ 0.13
$ & $ 2.4 $ & $ 676 $ & $ 1 $ & \tablenotemark{16} \\
$ 829.03 $ & $ 38.56 $ & C & $ 501 $ & $ 77 $ & $ 25.2 $ & $ 4.7
$ & $ 11 $ & $ 26 $ & $ 3 $ & \tablenotemark{7,17}~~~Kepler53c \\
$ 841.01 $ & $ 15.34 $ & C & $ 780 $ & \nodata & $ 14.8 $ & $ 1.1
$ & $ 4.6 $ & $ 55 $ & $ 2 $ & \tablenotemark{2,6}~~Kepler27b \\
$ 841.02 $ & $ 31.33 $ & C & $ 630 $ & $ 120 $ & $ 17.6 $ & $ 2.9
$ & $ 9.1 $ & $ 31 $ & $ 2 $ & \tablenotemark{6}Kepler27c \\
$ 869.02 $ & $ 36.28 $ & C & $ 600 $ & $ 140 $ & $ 65.2 $ & $ 7.7
$ & $ 27 $ & $ 26 $ & $ 4 $ & \tablenotemark{17}  \\
 \hline
$ 870.01 $ & $ 5.91 $ & C & $ 231 $ & $ 23 $ & $ 8.74 $ & $ 1 $ &
$ 7.9 $ & $ 149 $ & $ 2 $ & \tablenotemark{6}Kepler28b \\
$ 870.02 $ & $ 8.99 $ & C & $ 230 $ & $ 21 $ & $ 12.3 $ & $ 1.9 $
& $ 13 $ & $ 97 $ & $ 2 $ & \tablenotemark{6}Kepler28c \\
$ 872.01 $ & $ 33.60 $ & C & $ 191.1 $ & $ 9.2 $ & $ 54.6 $ & $
5.3 $ & $ 19 $ & $ 28 $ & $ 2 $ &
\tablenotemark{2,11}~~~Kepler46b \\
$ 880.01 $ & $ 26.44 $ & C & $ 860 $ & \nodata & $ 20.5 $ & $ 1.8
$ & $ 4.5 $ & $ 31 $ & $ 4 $ & \tablenotemark{17}  \\
$ 880.02 $ & $ 51.53 $ & P & \nodata & \nodata & \nodata &
\nodata & $ 13 $ & $ 18 $ & $ 4 $ & \tablenotemark{17}  \\
$ 884.02 $ & $ 20.49 $ & C & $ 837 $ & $ 98 $ & $ 175 $ & $ 3.7 $
& $ 14 $ & $ 36 $ & $ 3 $ & \tablenotemark{1,2}~~~ \\
$ 886.01 $ & $ 8.01 $ & C & $ 860 $ & \nodata & $ 63.1 $ & $ 2.1
$ & $ 12 $ & $ 105 $ & $ 3 $ & \tablenotemark{2,7}~~Kepler54b \\
$ 886.02 $ & $ 12.07 $ & C & $ 800 $ & \nodata & $ 98 $ & $ 12 $ & $ 53 $ & $ 55 $ & $ 3 $ & \tablenotemark{7}~~~Kepler54c \\ 
$ 902.01 $ & $ 83.92 $ & P & \nodata & \nodata & \nodata &
\nodata & $ 1.2 $ & $ 13 $ & $ 1 $ &  \\
$ 904.02 $ & $ 27.96 $ & P & \nodata & \nodata & \nodata &
\nodata & $ 18 $ & $ 35 $ & $ 5 $ & \tablenotemark{7}Kepler55b \\
\hline
$ 904.03 $ & $ 42.15 $ & P & \nodata & \nodata & \nodata &
\nodata & $ 13 $ & $ 15 $ & $ 5 $ & \tablenotemark{7}Kepler55c \\
$ 918.01 $ & $ 39.64 $ & C & $ 950 $ & \nodata & $ 8.25 $ & $
0.78 $ & $ 2.6 $ & $ 22 $ & $ 1 $ & \tablenotemark{2}~~~ \\
$ 928.01 $ & $ 2.49 $ & C & $ 120 $ & \nodata & $ 30.4 $ & $ 1.5
$ & $ 17 $ & $ 273 $ & $ 1 $ & \tablenotemark{1,20}~~~ \\
$ \tablenotemark{*} 935.01 $ & $ 20.86 $ & C & $ 1000 $ & \nodata
& $ 25.3 $ & $
1.9 $ & $ 5.8 $ & $ 44 $ & $ 4 $ &
\tablenotemark{1,10}~~~Kepler31b \\
$ 984.01 $ & $ 4.29 $ & C & $ 495 $ & $ 25 $ & $ 45.55 $ & $ 0.37
$ & $ 3.6 $ & $ 196 $ & $ 1 $ & \tablenotemark{2}~~~ \\
$ 989.03 $ & $ 16.16 $ & P & \nodata & \nodata & \nodata &
\nodata & $ 22 $ & $ 46 $ & $ 1 $ &  \\
$ 1061.01 $ & $ 41.81 $ & P & \nodata & \nodata & \nodata &
\nodata & $ 17 $ & $ 19 $ & $ 1 $ &  \\
$ 1081.01 $ & $ 9.96 $ & C & $ 1000 $ & \nodata & $ 74.2 $ & $
4.6 $ & $ 14 $ & $ 76 $ & $ 1 $ & \tablenotemark{2}~~~ \\
$ 1102.01 $ & $ 12.33 $ & C & $ 421 $ & $ 54 $ & $ 50.8 $ & $ 4.2
$ & $ 21 $ & $ 58 $ & $ 2 $ & \tablenotemark{2,5}~~Kepler24c \\
$ 1102.02 $ & $ 8.15 $ & C & $ 434 $ & $ 62 $ & $ 29.7 $ & $ 4.2
$ & $ 23 $ & $ 92 $ & $ 2 $ & \tablenotemark{2,5}~~Kepler24b \\

 \hline

\end{tabular}
 \end{table}

\addtocounter{table}{-1}

\begin{table} \footnotesize \caption{ - continued}
\begin{tabular}{|rr|lrrrrr|r|r|l|}
\hline \hline

KOI & Period\tablenotemark{a} & Model\tablenotemark{b}  & Period\tablenotemark{c}   & $\sigma_P\tablenotemark{d} $ & Amp\tablenotemark{e}   &$\sigma_A\tablenotemark{f} $&
Res\tablenotemark{g}  & $N$\tablenotemark{h}  &   Multi-  &  Ref.\tablenotemark{j} \\
     &  [d] &    & [d]      & [d]    &  [min]    &   [min]   &
[min] &  & plicity\tablenotemark{i}  & \\

 \hline
$ 1145.01 $ & $ 30.59 $ & C & $ 950 $ & \nodata & $ 76.8 $ & $
4.5 $ & $ 9.6 $ & $ 29 $ & $ 1 $ & \tablenotemark{2}~~~ \\
$ 1236.01 $ & $ 35.74 $ & P & \nodata & \nodata & \nodata &
\nodata & $ 22 $ & $ 25 $ & $ 2 $ &  \\
$ 1241.01 $ & $ 21.41 $ & C & $ 524 $ & $ 90 $ & $ 42 $ & $ 11 $
& $ 32 $ & $ 41 $ & $ 2 $ & \tablenotemark{7}Kepler56c \\
$ 1241.02 $ & $ 10.50 $ & C & $ 509 $ & $ 78 $ & $ 161 $ & $ 21 $
& $ 100 $ & $ 70 $ & $ 2 $ & \tablenotemark{1,7}~~Kepler56b \\
$ 1270.02 $ & $ 11.61 $ & C & $ 459 $ & $ 75 $ & $ 34.5 $ & $ 3.3
$ & $ 14 $ & $ 61 $ & $ 2 $ &
\tablenotemark{2,7,17}~~~~~Kepler57c \\
$ 1271.01 $ & $ 161.86 $ & P & \nodata & \nodata & \nodata &
\nodata & $ 86 $ & $ 6 $ & $ 1 $ & \tablenotemark{2}~~~ \\
$ \tablenotemark{*} 1285.01 $ & $ 0.94 $ & \nodata & \nodata &
\nodata & \nodata &
\nodata & \nodata & $ 998 $ & $ 1 $ & \tablenotemark{2,16}~~~ \\
$ 1353.01 $ & $ 125.87 $ & P & \nodata & \nodata & \nodata &
\nodata & $ 0.51 $ & $ 7 $ & $ 2 $ &  \\
$ 1426.01 $ & $ 38.87 $ & P & \nodata & \nodata & \nodata &
\nodata & $ 17 $ & $ 23 $ & $ 3 $ &  \\
$ 1426.02 $ & $ 74.92 $ & C & $ 970 $ & \nodata & $ 36.3 $ & $
4.2 $ & $ 6.8 $ & $ 10 $ & $ 3 $ &  \\
 \hline
$ 1426.03 $ & $ 150.02 $ & C & $ 820 $ & \nodata & $ 22.4 $ & $
2.3 $ & $ 1.3 $ & $ 7 $ & $ 3 $ &  \\
$ 1429.01 $ & $ 205.92 $ & P & \nodata & \nodata & \nodata &
\nodata & $ 3.5 $ & $ 5 $ & $ 1 $ &  \\
$ \tablenotemark{*} 1452.01 $ & $ 1.15 $ & \nodata & \nodata &
\nodata & \nodata &
\nodata & \nodata & $ 801 $ & $ 1 $ & \tablenotemark{16} \\
$ 1459.01 $ & $ 0.69 $ & P & \nodata & \nodata & \nodata &
\nodata & $ 2.8 $ & $ 493 $ & $ 1 $ &  \\
$ \tablenotemark{*}1474.01 $ & $ 69.73 $ & \nodata & \nodata & \nodata & \nodata & \nodata & \nodata & $ 14 $ & $ 1 $ & \tablenotemark{2}~~~ \\ 
$ 1529.01 $ & $ 17.98 $ & C & $ 520 $ & $ 76 $ & $ 61.7 $ & $ 7.4 $ & $ 27 $ & $ 38 $ & $ 2 $ & \tablenotemark{2,7}~~Kepler59c \\ 
$ \tablenotemark{*}1540.01 $ & $ 1.21 $ & \nodata & \nodata & \nodata & \nodata & \nodata & \nodata & $ 741 $ & $ 1 $ & \tablenotemark{16} \\ 

$  \tablenotemark{*}1543.01 $ & $ 3.96 $ & \nodata & \nodata &
\nodata & \nodata &
\nodata & \nodata & $ 231 $ & $ 1 $ & \tablenotemark{16} \\
$ \tablenotemark{*} 1546.01 $ & $ 0.92 $ & \nodata & \nodata &
\nodata & \nodata &
\nodata & \nodata & $ 993 $ & $ 1 $ &  \tablenotemark{16}\\
$ 1573.01 $ & $ 24.81 $ & C & $ 990 $ & \nodata & $ 39 $ & $ 1.4
$ & $ 3.3 $ & $ 32 $ & $ 1 $ & \tablenotemark{2}\\
 \hline
$ 1581.01 $ & $ 29.54 $ & C & $ 1000 $ & \nodata & $ 90 $ & $ 13
$ & $ 26 $ & $ 27 $ & $ 1 $ & \tablenotemark{2}\\
$ 1582.01 $ & $ 186.40 $ & P & \nodata & \nodata & \nodata &
\nodata & $ 3.4 $ & $ 6 $ & $ 1 $ &  \\
$ 1589.02 $ & $ 12.88 $ & C & $ 268 $ & $ 24 $ & $ 37.4 $ & $ 4.9
$ & $ 24 $ & $ 65 $ & $ 5 $ & \tablenotemark{17} \\
$ 1599.01 $ & $ 20.41 $ & P & \nodata & \nodata & \nodata &
\nodata & $ 26 $ & $ 34 $ & $ 1 $ & \tablenotemark{2,18} \\
$ 1675.01 $ & $ 14.62 $ & C & $ 510 $ & $ 110 $ & $ 17.7 $ & $
2.8 $ & $ 11 $ & $ 50 $ & $ 1 $ &  \\
$ 1747.01 $ & $ 20.56 $ & C & $ 760 $ & \nodata & $ 14.3 $ & $
2.5 $ & $ 8 $ & $ 31 $ & $ 1 $ &  \\
$ 1751.02 $ & $ 21.00 $ & P & \nodata & \nodata & \nodata &
\nodata & $ 6.2 $ & $ 35 $ & $ 2 $ &  \\
$ 1781.01 $ & $ 7.83 $ & P & \nodata & \nodata & \nodata &
\nodata & $ 1.4 $ & $ 86 $ & $ 2 $ &  \\
$ 1802.01 $ & $ 5.25 $ & C & $ 232 $ & $ 21 $ & $ 6.91 $ & $ 0.61
$ & $ 5.3 $ & $ 177 $ & $ 1 $ &  \\
$ 1805.01 $ & $ 6.94 $ & C & $ 226 $ & $ 25 $ & $ 3.72 $ & $ 0.58
$ & $ 4.4 $ & $ 135 $ & $ 3 $ &  \\

\hline
\end{tabular}
\label{tab:TTV_Interesting}
\end{table}
\addtocounter{table}{-1}

\begin{table} \footnotesize \caption{ - continued}
\begin{tabular}{|rr|lrrrrr|r|r|l|}
\hline \hline

KOI & Period\tablenotemark{a} & Model\tablenotemark{b}  & Period\tablenotemark{c}   & $\sigma_P\tablenotemark{d} $ & Amp\tablenotemark{e}   &$\sigma_A\tablenotemark{f} $&
Res\tablenotemark{g}  & $N$\tablenotemark{h}  &   Multi-  &  Ref.\tablenotemark{j} \\
     &  [d] &    & [d]      & [d]    &  [min]    &   [min]   &
[min] &  & plicity\tablenotemark{i}  & \\

\hline
$ 1840.01 $ & $ 7.04 $ & C & $ 1000 $ & \nodata & $ 31.5 $ & $
2.1 $ & $ 9.4 $ & $ 119 $ & $ 1 $ & \tablenotemark{2}\\
$ 1856.01 $ & $ 46.30 $ & C & $ 850 $ & \nodata & $ 42.7 $ & $
3.4 $ & $ 4.5 $ & $ 21 $ & $ 1 $ &  \\
$ 1884.01 $ & $ 23.12 $ & P & \nodata & \nodata & \nodata &
\nodata & $ 8.7 $ & $ 18 $ & $ 2 $ &  \\
$ 1973.01 $ & $ 3.29 $ & C & $ 417 $ & $ 81 $ & $ 19.1 $ & $ 1.7
$ & $ 13 $ & $ 166 $ & $ 1 $ &  \\
$ 1986.01 $ & $ 148.46 $ & C & $ 594 $ & $ 81 $ & $ 19.3 $ & $
5.1 $ & $ 4.5 $ & $ 7 $ & $ 1 $ &  \tablenotemark{18} \\
$ 2037.03 $ & $ 8.56 $ & C & $ 510 $ & $ 170 $ & $ 18.8 $ & $ 2.1
$ & $ 10 $ & $ 63 $ & $ 2 $ &  \\
$ 2038.01 $ & $ 8.31 $ & C & $ 1000 $ & \nodata & $ 39.5 $ & $
6.2 $ & $ 21 $ & $ 76 $ & $ 4 $ &  \\
$ 2038.02 $ & $ 12.51 $ & C & $ 680 $ & \nodata & $ 41.3 $ & $
5.3 $ & $ 27 $ & $ 71 $ & $ 4 $ &  \\
$ 2291.01 $ & $ 44.30 $ & P & \nodata & \nodata & \nodata &
\nodata & $ 11 $ & $ 18 $ & $ 1 $ &  \\
$ 2613.01 $ & $ 51.58 $ & P & \nodata & \nodata & \nodata &
\nodata & $ 19 $ & $ 14 $ & $ 1 $ &  \\

\hline
\end{tabular}

\tablecomments{
$^a$Orbital Period. 
$^b$Model type: C represents a cosine superimposed on a linear trend, P represents a parabolic fit, while '\nodata' means no fit. 
$^c$Best-fit period of the O-C data using model C. 
$^d$Period uncertainty. 
$^e$The amplitude of the cosine fit. 
$^f$Amplitude uncertainty. 
$^g$Residual scatter (1.483 times their MAD). 
$^h$Number of TT measurements. 
$^i$Number of detected planets in the system  (see B12). 
$^j$Reference.\\
$^*$Discussed in Section~\ref{comments}.\\
 $^1$F11. $^2$F12. $^3$\citet{ballard11}. $^4$\citet{cochran11}. 
$^5$\citet{ford12a}. 
$^6$\citet{steffen12a}.
$^7$\citet{steffen13}. 
$^8$\citet{carter12}. $^9$\citet{holman10}. 
$^{10}$\citet{fabrycky12}. 
$^{11}$\citet{nesvorny12}. 
$^{12}$\citet{nesvorny13}. $^{13}$\citet{wang12}. $^{14}$\citet{santerne12}. $^{15}$\citet{tingley11}. $^{16}$\citet{szabo13}. $^{17}$\citet{xie12}. $^{18}$\citet{ofir12}. $^{19}$\citet{ragozzine12}. 
$^{20}$\citet{steffen11}.
}
\end{table}




\newpage

\section{KOIs with short-period TTV}

\begin{figure*}
\centering
\resizebox{16cm}{11cm}
{\includegraphics{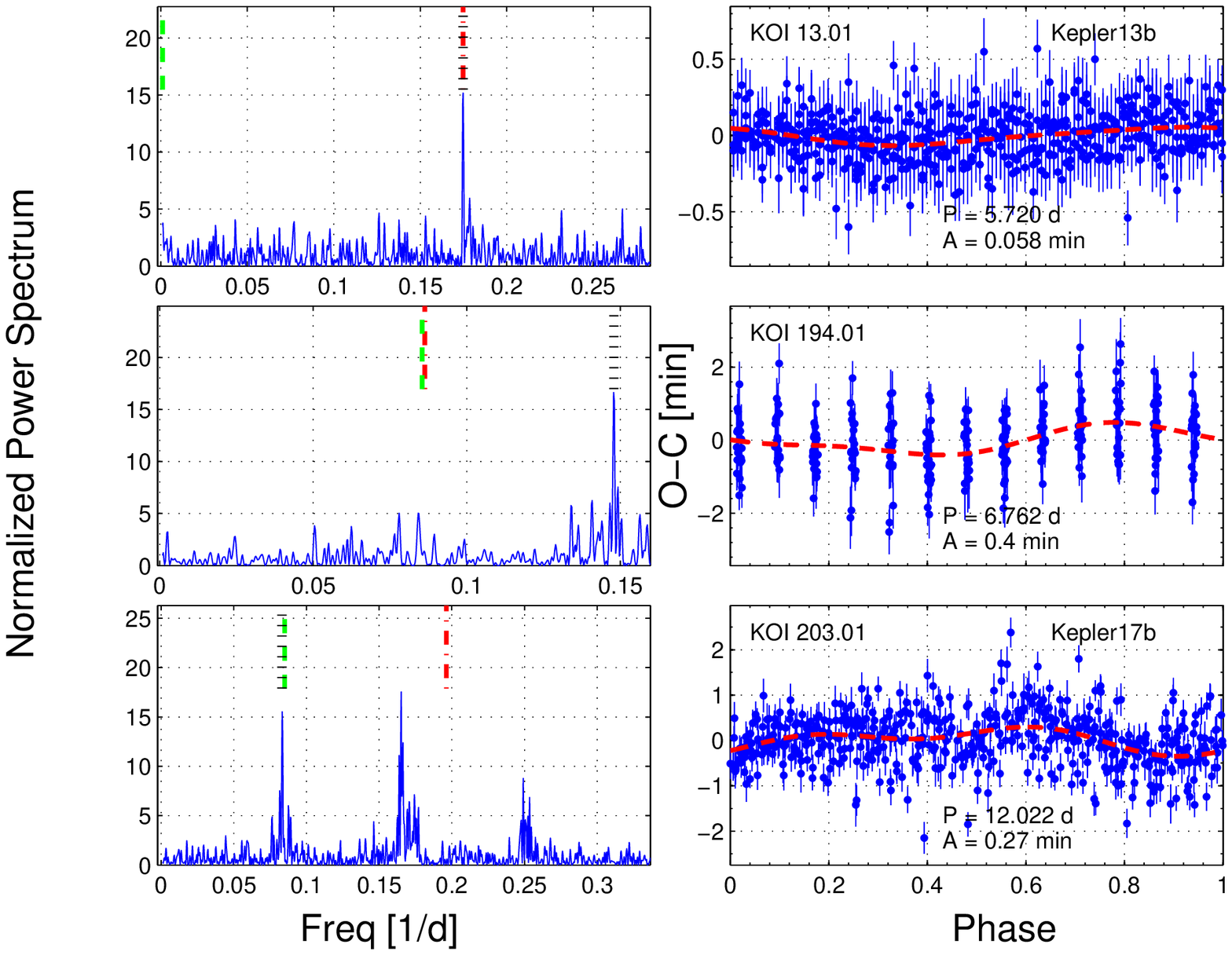}}
\caption{The KOIs with short-period TTVs.  For each KOI, the plot shows the LS
periodogram
and the phase-folded O-Cs. The dotted black line represents
the folding period, the dashed green line
the stellar activity frequency or one of its aliases, if
present in the stellar light curve, and the dash-dotted red
line  the frequency induced by the sampling. 
The phase-folded light-curve panels include a two-harmonic fit.
}
\label{TTV_short1}
\end{figure*}

\begin{figure*}
\centering
\resizebox{16cm}{11cm}
{\includegraphics{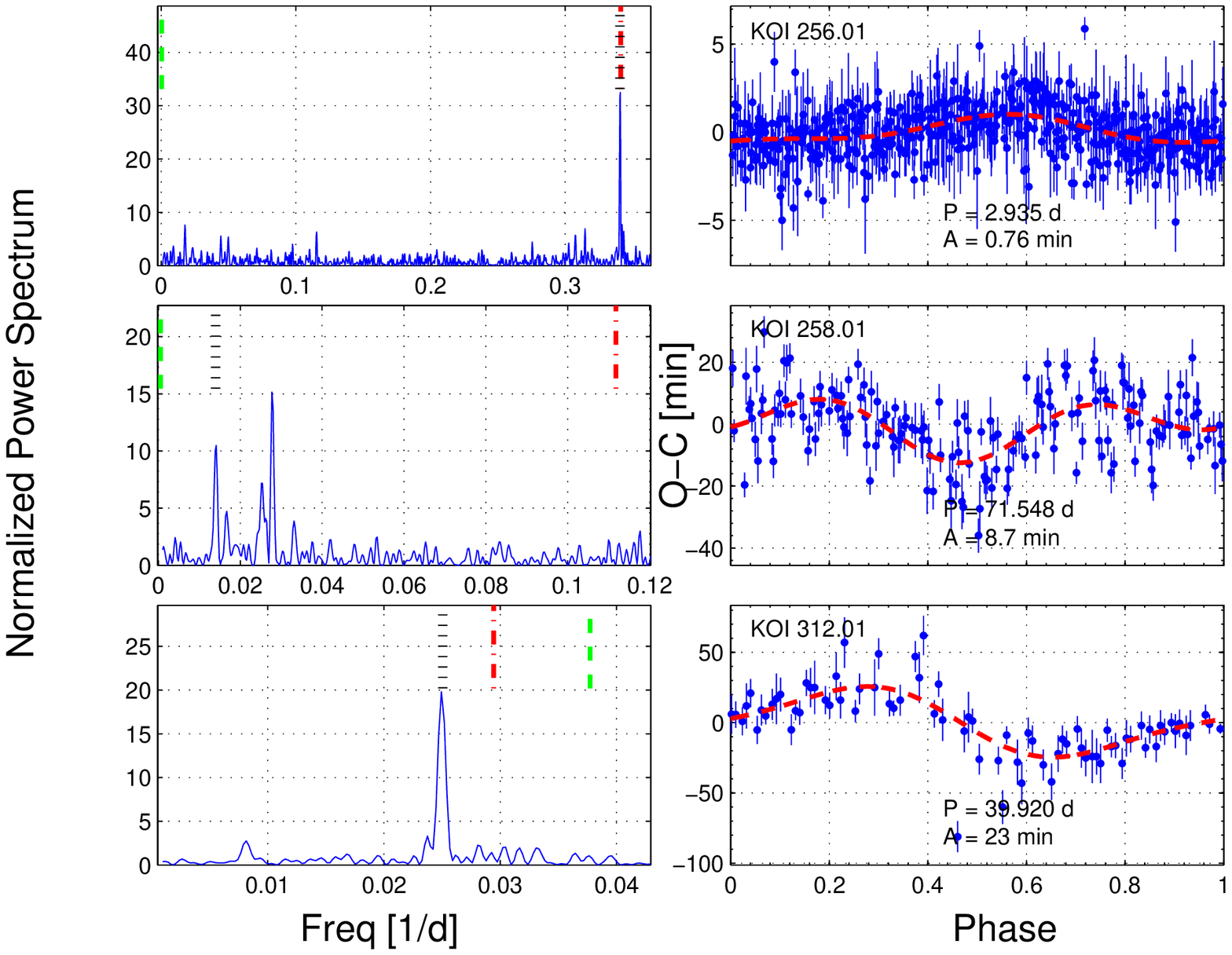}}
\caption{The KOIs with short-period TTVs.  For each KOI, the LS
periodogram
and the phase-folded O-Cs are plotted (see Figure~\ref{TTV_short1} for
details).
}
\label{TTV_short2}
\end{figure*}

\begin{figure*}
\centering
\resizebox{16cm}{11cm}
{\includegraphics{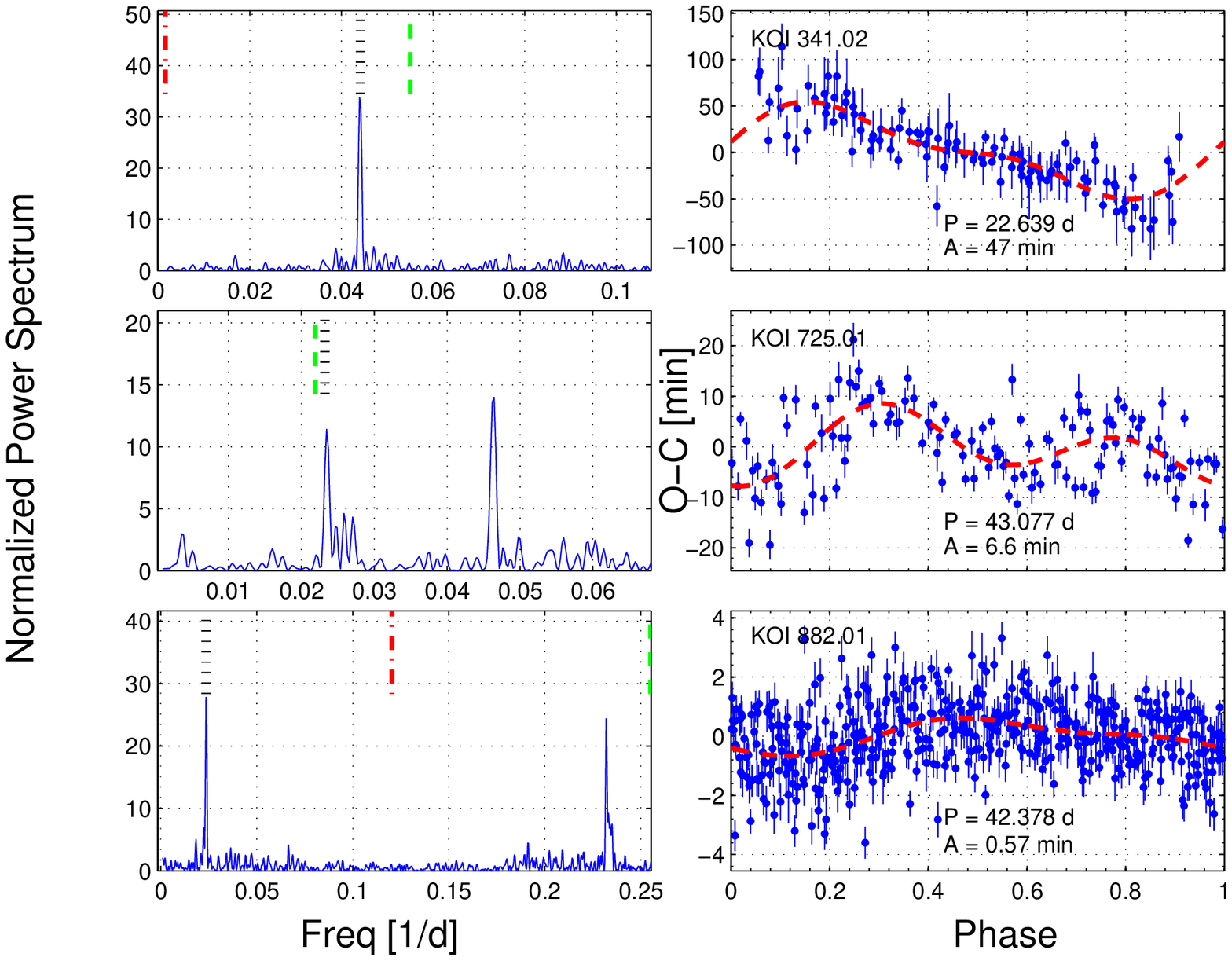}}
\caption{The KOIs with short-period TTVs.  For each KOI, the LS
periodogram
and the phase-folded O-Cs are plotted  (see Figure~\ref{TTV_short1} for
details).
}
\label{TTV_short3}
\end{figure*}

\begin{figure*}
\centering
\resizebox{16cm}{11cm}
{\includegraphics{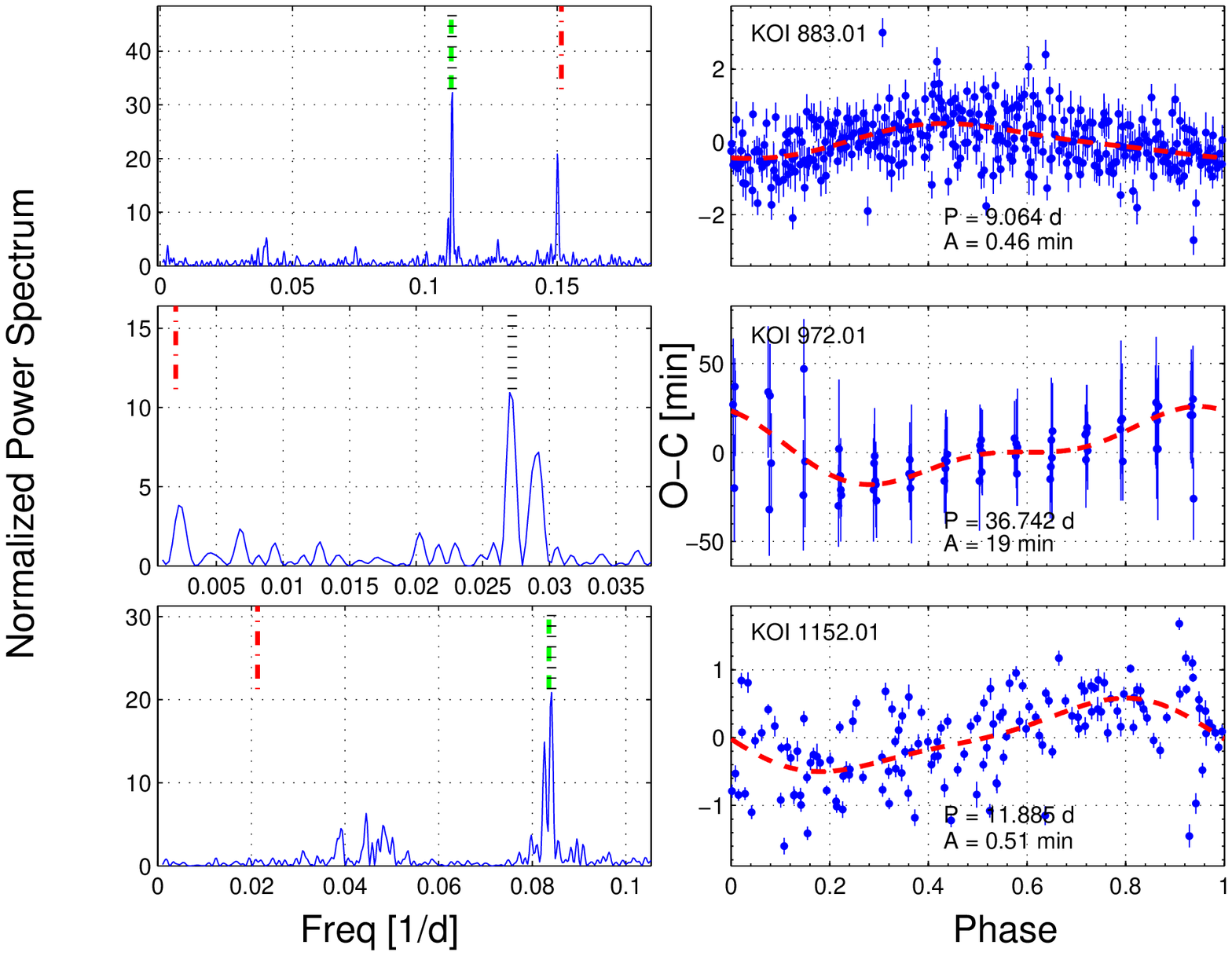}}
\caption{The KOIs with short-period TTVs.  For each KOI, the LS
periodogram
and the phase-folded O-Cs are plotted  (see Figure~\ref{TTV_short1} for
details).
}
\label{TTV_short4}
\end{figure*}

\begin{figure*}
\centering
\resizebox{16cm}{11cm}
{\includegraphics{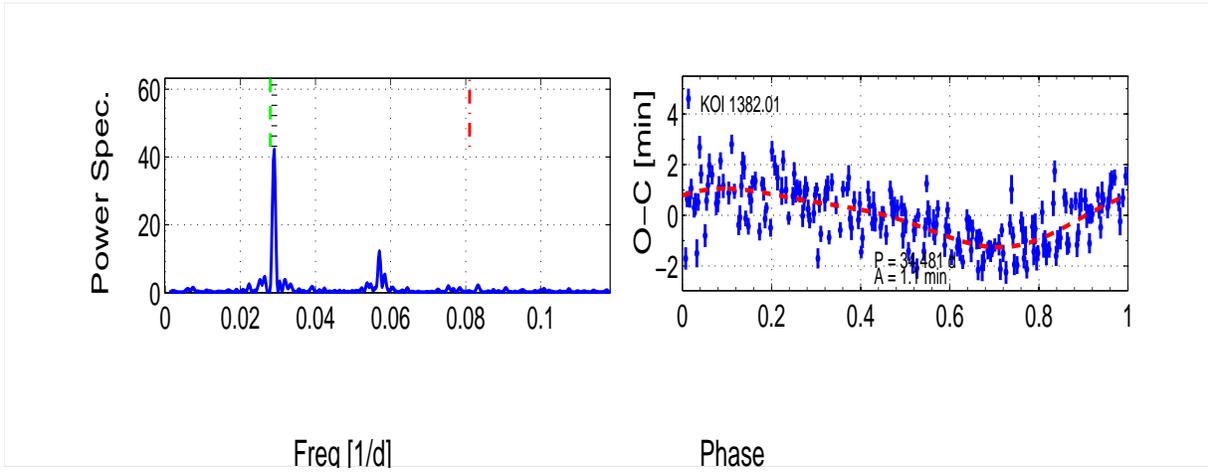}}
\caption{The KOIs with short-period TTVs.  For each KOI, the LS
periodogram
and the phase-folded O-Cs are plotted  (see Figure~\ref{TTV_short1} for
details).
}
\label{TTV_short5}
\end{figure*}

Our LS analysis of the O-Cs yielded also $13$ systems
with highly significant short-period TTV modulations, in the
range of 3 to 72
days. They were found using the same criterion as the one used in Section~\ref{long}
--- LS peak with
FAP lower
than 3$\times$10$^{-4}$.
The modulation
amplitudes were relatively small, in the range of $0.06- 46$ minutes,
and their detection was possible only because of the modulation
periodicity and the long time span of the data.
Figures~\ref{TTV_short1}--\ref{TTV_short5} show the LS
periodograms and the phase-folded O-Cs of the $13$ systems, 
where one can see the prominent
peaks of the
periodograms. Table~\ref{tab:short} lists the periods and
amplitudes found.
The table includes references to Section~\ref{comments},
where we briefly comment on these systems.

As pointed out by \citet[][hereafter Sz13]{szabo13},
not all detected short-period modulations are due to physical
TTVs. An {\it apparent} TTV periodicity can be induced either
 by the long-cadence sampling of {\it Kepler}, or by an interference with a periodic stellar activity.

The finite sampling rate of the observations may cause a shift in
the orbital phases of the observations during a transit, inducing
an apparent shift of the derived timing of that transit. This can
evoke a periodic TTV, with a period of

\begin{equation}
P_{\rm induced} = \frac
{P_{\rm orb}}
{
\frac{P_{\rm orb}}{P_{\rm samp}}-\lfloor \frac{P_{\rm orb}}{P_{\rm samp}}
\rfloor
} \ ,
\end{equation}
where $P_{\rm orb}$ is the orbital period, $P_{\rm samp}$
is the sampling cadence and $\lfloor x \rfloor$ is the floor of
$x$ (Sz13).
Note that the induced periodicity is observed with "sampling"
intervals equal to the planetary orbit, and therefore the relevant Nyquist
frequency is  $1/(2P_{\rm orb})$. If the induced frequency
 is larger, we will detect one of its aliases.
We found two cases which suggested that the O-Cs included such
effect (see below).

The other effect is due to the stellar spot activity, which
modulates the stellar intensity with the stellar rotational
period. Spot crossing \citep[e.g.,][]{sanchis12} during a
transit, or a slope of the stellar brightness during a
transit, can cause a shift in the derived transit timing,
inducing an apparent O-C periodicity with the stellar rotational
period. In fact, $12$ out of the $13$ systems with short periodic
modulation showed a high level of
stellar activity, and we had to check whether the detected TTV
periodicity was due to that activity. 

To find the frequency of the presumed sampling-induced
periodicity, we used for each KOI its $P_{\rm orb}$ from Table
\ref{tab:ephemeris}, and the pertinent $P_{\rm samp}$. This was
about 29.424
minutes for the long cadence, the exact value taken to be the median of the
differences of the observed timings of that KOI.
We searched for stellar spot periodicity using the autocorrelation technique
\citep[e.g.,][]{mcquillan13}, and, if present, checked whether its
frequency, or one of its aliases, was equal to the TTV frequency. We
mark the pertinent frequencies in
Figures~\ref{TTV_short1}--\ref{TTV_short5}.

\begin{table}[h!]
\footnotesize
\caption{KOIs with significant short-period TTVs }

\begin{tabular}{|rr|rrrrr|r|r|l|}

\hline \hline

KOI & Period\tablenotemark{a} & Period\tablenotemark{b}   & $\sigma_P\tablenotemark{c} $ & Amp\tablenotemark{d}   &$\sigma_A\tablenotemark{e} $&
Res\tablenotemark{f}  & $N$\tablenotemark{g}  &   Multi-  &  Ref.\tablenotemark{i} \\
     &  [d]  & [d]      & [d]    &  [min]    &   [min]   &
[min] &  & plicity\tablenotemark{h}  & \\

\hline

$ \tablenotemark{*}13.01 $ & $ 1.76 $  & $ 5.72 $ & $ 0.015 $

& $ 0.0578 $ & $

0.0094 $ & $ 0.15 $ & $ 520 $ & $ 1 $ &

\tablenotemark{1}Kepler13b \\

$  \tablenotemark{*}194.01 $ & $ 3.12 $ & $ 6.762 $ & $ 0.016

$ & $ 0.405 $ & $

0.06 $ & $ 0.71 $ & $ 293 $ & $ 1 $ &  \\

$ \tablenotemark{*} 203.01 $ & $ 1.49 $ & $ 12.022 $ & $

0.051 $ & $ 0.271 $ &

$ 0.034 $ & $ 0.49 $ & $ 505 $ & $ 1 $ &

\tablenotemark{2,3,4}~~~~Kepler17b \\

\hline

$ \tablenotemark{*} 256.01 $ & $ 1.38 $  & $ 2.9353 $ & $

0.0043 $ & $ 0.765 $ &

$ 0.077 $ & $ 1.2 $ & $ 519 $ & $ 1 $ &   \tablenotemark{3} \\

$  \tablenotemark{*}258.01 $ & $ 4.16 $ & $ 71.5 $ & $ 1.5 $

& $ 8.75 $ & $

0.79 $ & $ 9.1 $ & $ 174 $ & $ 1 $ &  \\

$  \tablenotemark{*}312.01 $ & $ 11.58 $  & $ 39.92 $ & $ 0.48

$ & $ 23.4 $ & $

2 $ & $ 11 $ & $ 78 $ & $ 2 $ &  \\

\hline

$ 341.02 $ & $ 4.70 $  & $ 22.64 $ & $ 0.15 $ & $ 46.5 $ & $

2.7 $ & $ 17 $ & $ 112 $ & $ 2 $ &  \\

$ \tablenotemark{*} 725.01 $ & $ 7.30 $  & $ 43.08 $ & $ 0.56

$ & $ 6.59 $ & $

0.59 $ & $ 5.8 $ & $ 132 $ & $ 1 $ &  \\

$  \tablenotemark{*}882.01 $ & $ 1.96 $ & $ 42.38 $ & $ 0.63

$ & $ 0.571 $ & $

0.068 $ & $ 1.1 $ & $ 484 $ & $ 1 $ &    \tablenotemark{3} \\

\hline

$  \tablenotemark{*}883.01 $ & $ 2.69 $ & $ 9.064 $ & $ 0.045

$ & $ 0.457 $ & $

0.048 $ & $ 0.58 $ & $ 354 $ & $ 1 $ &   \tablenotemark{3} \\

$ 972.01 $ & $ 13.12 $  & $ 36.74 $ & $ 0.34 $ & $ 19 $ & $

1.9 $ & $ 9.8 $ & $ 69 $ & $ 1 $ &  \\

$ \tablenotemark{*} 1152.01 $ & $ 4.72 $  & $ 11.885 $ & $

0.049 $ & $ 0.509 $ &

$ 0.063 $ & $ 0.43 $ & $ 148 $ & $ 1 $ &   \tablenotemark{3} \\

\hline

$  \tablenotemark{*}1382.01 $ & $ 4.20 $ & $ 34.48 $ & $ 0.66

$ & $ 1.096 $ & $

0.079 $ & $ 0.8 $ & $ 201 $ & $ 1 $ &   \tablenotemark{3} \\

\hline

\end{tabular}
\tablecomments{
$^a$Orbital Period. 
$^b$Best-fit period of the O-C data. 
$^c$Period uncertainty. 
$^d$The amplitude of the cosine model. $^e$Amplitude uncertainty. 
$^f$Residual scatter (1.483 times their MAD). 
$^g$Number of TT measurements. $^h$Number of planets in the system according to B12. $^i$Reference.\\
$^*$Discussed in Section~\ref{comments}.\\
 $^{1}$\citet{shporer11}.  $^{2}$\citet{desert11}.  $^{3}$\citet{szabo13}.  $^{4}$\citet{bonomo12}.
}

 \label{tab:short}
\end{table}

\newpage

\section{Comments on Individual Systems}
\label{comments}

In this section we comment on a few KOIs from Tables
\ref{tab:TTV_Interesting}--\ref{tab:short}.
In particular, we phase-folded the light curves of all $143$ systems
with their orbital period, and searched for a secondary dip.
For eleven systems we found a significant secondary dip, in
most cases at phase $\sim0.5$, which we interpreted as either an
eclipse of a secondary star, or a planetary occultation. 
We also point out any periodic TTV modulation that could have been
induced either by the long cadence sampling or by the stellar
spot periodic activity. Some of these systems were analyzed in a
similar way by Sz13, who used six quarters of {\it Kepler} data
to look for TTV periodicity.

\begin{itemize}

\item
KOI-$13.01$ (Figure 17):
The O-Cs LS periodogram displays a prominent peak, corresponding
to the induced sampling frequency. The folded light curve
displays a shallow occultation.

\item
KOI-$142.01$ (Figure 4):
The TTV modulation has one of the largest amplitudes in the sample. One cosine function was not enough to model the modulation, and therefoe the O-Cs include at least two different
frequencies. This might be the result of some non-linear effect of the dynamical
interaction. \citet{nesvorny13} derived the parameters of the unseen planet causing this TTV.

\item
KOI-$190.01$ (Figure 5): 
This system is probably an eclipsing binary (=EB) orbiting a third distant star, causing the light time travel (LITE) effect \citep{santerne12}.

\item
KOI-$194.01$ (Figure 17):  The folded light curve displays a
shallow occultation.

\item
KOI-$203.01$ (Figure 17):
The TTV LS periodogram displays two prominent peaks. The higher
frequency is the
first harmonic of the other. The lower-peak frequency coincides with the stellar rotation, which has a modulation with a period of  $12.05$ day. Sz13 reached the same conclusion.

\item
KOI-$256.01$ (Figure 18):
The O-C LS periodogram displays a prominent peak, corresponding
to the induced sampling frequency. Sz13 found a $41.8$ day period in the TTV.

\item
KOI-$258.01$ (Figure 18):
The folded light curve reveals a dip around phase $0.5$,
probably a secondary eclipse.  Therefore the system is probably an EB.
The star has a significantly high level of activity, probably due
to stellar pulsations. The O-C LS periodogram displays two prominent peaks. The higher frequency is the first harmonic of the other.

\item
KOI-$312.01$ (Figure 18):
The autocorrelation of the stellar photometry reveals a weak but
stable modulation with a short period of $0.17073$ day. The green line in the figure is an alias of this frequency.

\item
KOI-$341.01$ :
The orbital period used in our analysis is half the one published in B12. This KOI does not have a significant TTV and we do  not include it  in our tables.

\item
KOI-$609.01$ (Figure 9):
The folded light curve displays a shallow occultation. \citet{santerne12} found it to be an EB.

\item
KOI-$725.01$ (Figure 19):
The folded light curve displays a shallow occultation.
The stellar photometry shows pulsations with a period of $8.58$ day.
The O-C LS periodogram displays two prominent peaks.
The higher frequency is the
first harmonic of the other.

\item
KOI-$823.01$ (Figure 10):
The folded light curve displays a relatively deep occultation. Sz13 found it
to be a multi-periodic candidate.

\item
KOI-$882.01$ (Figure 20):
The photometry displays strong stellar pulsations with a
frequency of $3.921$ day, very close to twice the orbital period.
The second peak in the periodogram is an alias of the first one,
relative to the pulsation Nyquist frequency. Sz13 found the same
TTV periodicity, with a noisier periodogram.

\item
KOI-$883.01$ (Figure 20):
The O-C LS periodogram displays two prominent peaks. The higher
one coincides with the stellar rotation, with a period of $9.02$
day, and the smaller one with the induced sampling frequency. Sz13
reached the same conclusion.

\item
KOI-$928.01$ (Figure 12):
The orbital period is probably twice the one published by B12.
This system is probably an EB orbiting a third distant star,
causing a LITE effect \citep{steffen11}

\item
KOI-$935.01$ (Figure 12):
The folded light curve probably displays an occultation.

\item
KOI-$984.01$ (Figure 12):
O-Cs started to deviate from the strictly cosine function at
BJD$\sim2454900+ 100$.


\item
KOI-$1152.01$ (Figure 16):
The O-C LS periodogram displays one prominent peak, with
a frequency very close to the stellar rotational one, at a period
of 2.95 day. Sz13 found a TTV period which was twice the period we
found, and did not associate it with the stellar modulation.
They also detected a secondary eclipse not in phase 0.5, and
concluded that the system is an eccentric EB.

\item
KOI-$1285.01$ (Figure 13):
The folded light curve, with the orbital period of $0.9374$ day,
reveals a dip at about phase $0.5$, probably a secondary eclipse.
Therefore the system is probably an EB.
The O-Cs displays coherent modulations, but not a clear stable
periodicity.
The star has a significantly high level of periodic activity, with
a period of 0.9362 day, which is close to but not identical with the
orbital period.
Sz13 identified a few different possible TTV periods. They
suspected that two of their periods were affected by the stellar
modulation.

\item
KOI-$1382.01$ (Figure 21):
The folded light curve reveals a dip at about phase $0.5$,
probably
a secondary eclipse. Therefore the system is probably an EB.
The O-C LS periodogram displays one prominent peak, with a
frequency that coincides with one of the aliases of the stellar
rotational one, at a period of 4.79 day.
Sz13 identified the same TTV period, although with a much stronger
first harmonic. They failed to notice that the TTV periodicity
was the result of the stellar rotation.

\item
KOI-$1452.01$ (Figure 14):
The folded light curve, with an orbital period of $1.1522$ day,
reveals a dip at about phase $0.5$, probably a secondary eclipse,
and therefore the system is probably an EB.
The O-Cs display a coherent modulation, but not a clear stable
periodicity.
The stellar photometry displays strong stellar pulsations with
frequencies of 0.65597, 0.7097 and 0.83 day$^{-1}$. Sz13 found it
to be a multi-periodic candidate.

\item
KOI-$1474.01$ (Figure 14):
The TTV looks significant with a period of about $400$ day, but the shape of the modulation is uncommon. 

\item
KOI-$1540.01$ (Figure 14):
This is a grazing EB with a period of  2.4158 day, twice the period of B12.
Sz13 had an extensive discussion on this system, but did not notice the correct period.

\item
KOI-$1543.01$ (Figure 14):
The folded light curve, with the orbital period of $ 3.9643$ day,
reveals a dip at about  phase $0.5$, probably a secondary eclipse.
Therefore the system is
 probably an EB.
The stellar photometry displays a strong periodicity of $4.03$ day, probably due to stellar rotation. Sz13 found a $97$-day period in the O-Cs, suggesting it was a false positive.

\item
KOI-$1546.01$ (Figure 14):
The folded light curve, with the orbital period of $0.9176$ day,
reveals a dip at about phase $0.5$, probably a secondary eclipse.
Therefore the system is
 probably an EB.
The stellar photometry displays a strong periodicity of $0.933$
day, probably due to stellar rotation, a period very close to but not
identical with the orbital period. Sz13 found it to be a
multi-periodic variable.

\end{itemize}

\section{Discussion}
\label{discussion}

We present here $143$ KOIs with highly significant TTVs,
$130$ with long-term modulations (Section 4, Table 4), and 13
KOIs with short-period low-amplitude TTV periodicities (Section
5, Table 5). Out of the $130$ systems,
$85$ show clear periodicities, with well
determined periods and amplitudes. Another $39$ KOIs have
periods too long to be established without a doubt.
For those we need to wait for more data
before the TTV period can be safely determined.
Another six systems display coherent modulations, but not a
clear stable periodicity. 

\begin{figure*}
\centering
\resizebox{11cm}{9cm}
{\includegraphics{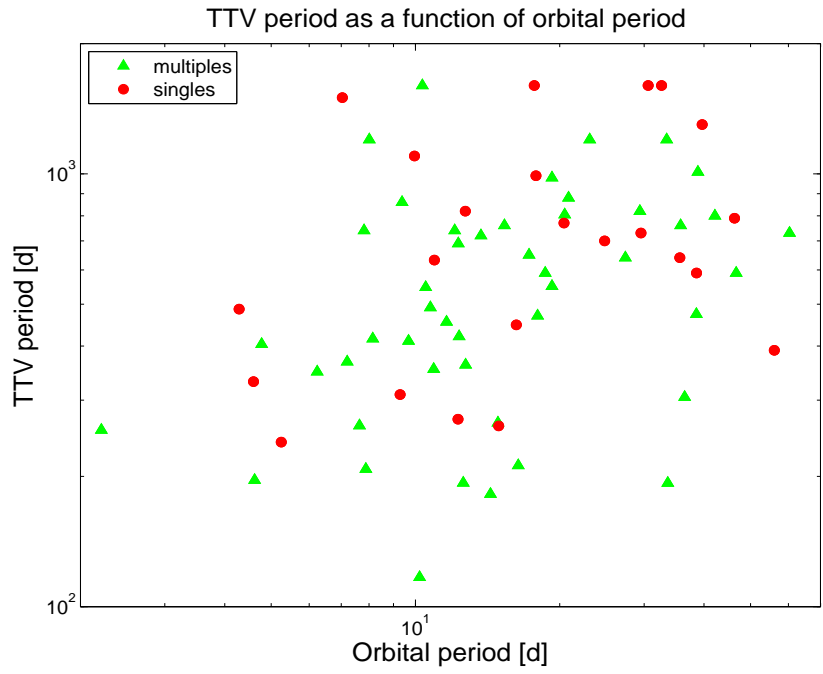}}
\caption{The TTV period as a function of the orbital period. 
Red circles represent single KOIs, and green triangles represent
multiples.
}
\label{PvsOrbitalP}
\end{figure*}

\begin{figure*}
\centering
\resizebox{11cm}{9cm}
{\includegraphics{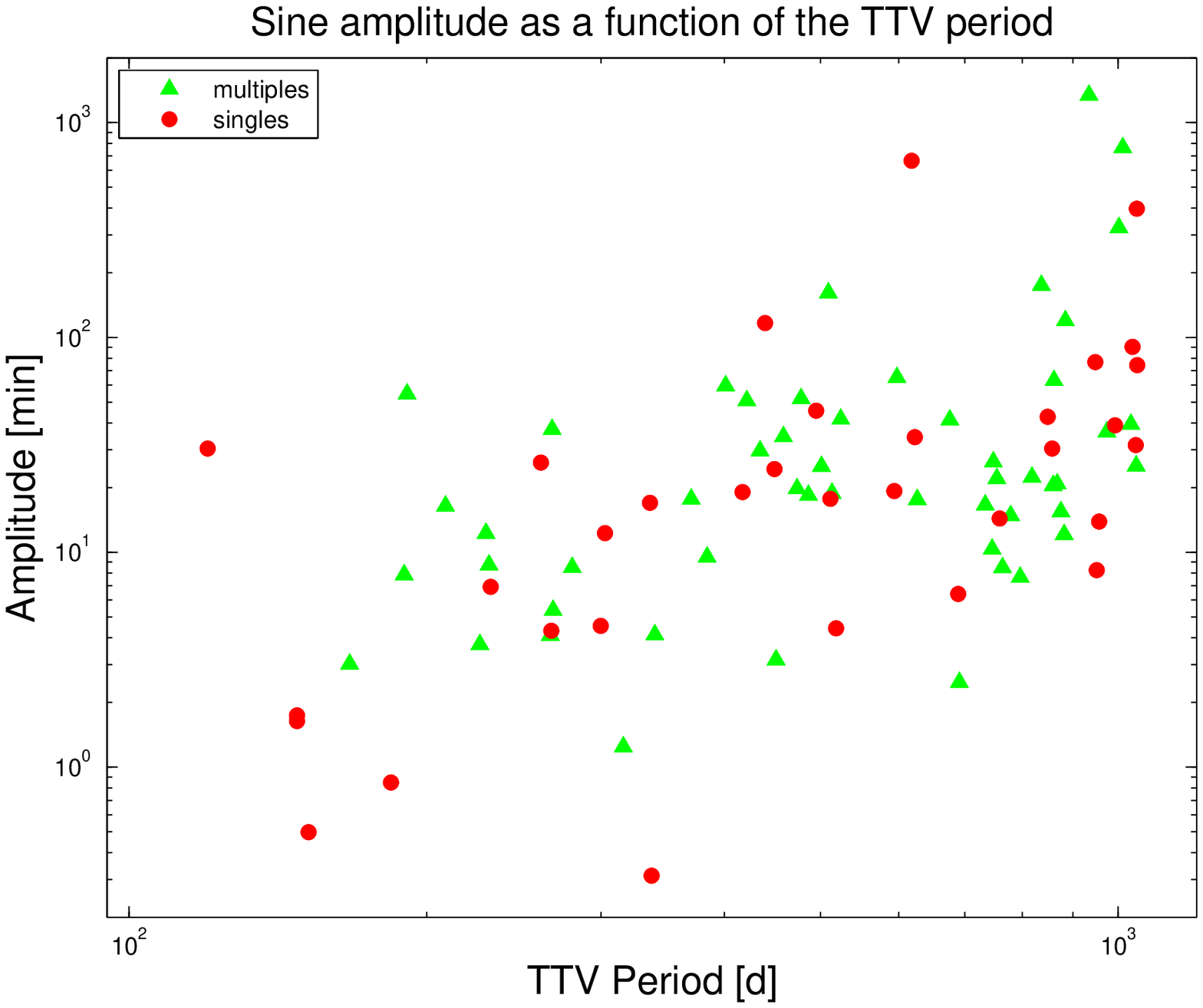}}
\caption{The amplitude of the TTV modulation as a function of its
period.
Red circles represent single KOIs, and green triangles represent
multiples.
}
\label{AvsP}
\end{figure*}

We have found an indication for some correlation, of $0.48$, 
between the KOI period
and the period of its TTV, as can be seen in
Figure~\ref{PvsOrbitalP}. This is of no surprise, as the orbital
period of a planet determines the natural time scale of the dynamical
interaction, and therefore
one can expect the TTV periodicity to be correlated with this time scale. 
Another correlation, of $0.51$, between the amplitudes and the
periods of the
detected TTV periodicities, emerged from our sample (see
Figure~\ref{AvsP}), as was predicted, for example, by
\citet{agol05}. The same correlation appeared when we plotted the
amplitude in units of the KOI orbital period.

We point out a possible non-dynamical origin of some of the TTVs
presented here. In particular, the short-period modulations could
be due to either the long cadence sampling of {\it Kepler} or
the stellar spot periodic activity (Sz13). We found evidence that
five out of the $13$ short-period detected TTVs are due to the
stellar periodicity.
We also found that KOI-$13.01$ and $883.01$ show a periodicity
induced by the Kepler sampling.

The sample of $143$ KOIs with significant TTVs includes $60$
systems discussed by F11 ($18$ KOIs), F12 ($38$ KOIs),
\citet{steffen12a}
($8$ KOIs), and \citet{steffen13} ($15$ KOIs),  all
based only on a fraction of the data available
now. References to those four works can be found in
Table~\ref{tab:TTV_Interesting}.
It is interesting to compare the analysis of F11,
F12 and  \citet[]{steffen12b,steffen13}  on one hand and the
present results on the other
hand,
and see how doubling the time span can change our assessment of
the nature of the modulation. In many cases the time span of the
first six quarters was not long enough to detect a local maximum
{\it and} minimum of the TTV modulation, and therefore the
periodicity of the modulation could not be estimated.  One
illustrious example is KOI-142, with its peak-to-peak amplitude
of more than 1200 min \citep[see][]{nesvorny13}.

One could hope that the accumulating details of the observed TTV
could give some hints for the orbital elements of the perturbing
unseen
planet, at least for some of the single KOIs. However,
as discussed already by \citet{holman05} and \citet{agol05}, the
amplitude and
periodicity of the TTV modulation depends on various parameters,
in particular the mass and the orbital period of the unseen planet and
how close the orbits of the two planets are to some mean motion resonance
\citep[e.g.,][]{lithwick12}.
Therefore,
it is quite difficult to deduce the parameters of the unseen
planet,
although some stringent constraints can be derived, as was done
by \citet{ballard11} and \citet{nesvorny12,nesvorny13}. We hope
that the available catalog will
motivate a similar work on other single-KOI systems with
significant TTV.

One parameter that has interesting implications on our
understanding of planetary formation is the relative inclination between the
orbital plane of the observed planet and that of the presumed
interacting planet
for the cases of single KOIs with significant TTVs. Relative inclination can
induce a precession of the orbital motion of the observed planet, which
can manifest
itself in a modulation of the transit duration and depth.
Although the focus of the present work is on the TTVs, the
catalog, which includes derived TDVs and TPVs, can, in principle, help to identify
systems with a relative inclination. 
Furthermore, stringent upper limits on TDVs
and TPVs for systems with detected TTVs can help to constrain the relative
inclinations between the planets.

However, an observed precession is not necessarily induced by a
planet with a non-vanishing relative inclination. An observed
precession of the orbital plane
just proves that the total angular momentum of the system is not
parallel to the orbital angular momentum of the transiting
planet.
The origin of the precession could also be a misalignment of the
stellar rotation axis relative to the angular momentum of the
planet, an idea that was unthinkable not long ago, but has now
a solid evidence in the accumulating data  \citep[see][]{winn11}.
An example is KOI-13 \citep[][see also
Figure~\ref{KOI13_TDV}]{szabo11}. Regardless, we suggest that
systems with detected significant TDVs and TPVs deserve further
close study.

Finally, we present here a systematic TTV analysis of twelve
quarters of {\it Kepler} observations of all KOIs.  One could
expect that the derived TTVs, for the single KOIs in particular,
could help in constructing a {\it statistical} picture of the
frequency and
architecture of the population of the planetary multiple systems
of the Kepler KOIs \citep[e.g.,][]{ford11, ford12b, lissauer11b,steffen10}. 
To perform such a
statistical analysis one needs to model the dependence of the
detectability of TTV coherent modulation on the parameters of the
unseen perturbing planet.
The present catalog can
be used for such a study.

\acknowledgments 
We thank the referee for his/her extremely valuable
remarks and suggestions.
The research leading to these results has
received funding from the European Research Council under the
EU's
Seventh Framework Programme (FP7/(2007-2013)/ ERC Grant Agreement
No.~291352) and from the ISRAEL SCIENCE FOUNDATION (grant
No.~1423/11).
All photometric data presented in this paper were
obtained from the Mikulsky Archive for Space Telescopes (MAST).
STScI is operated by the Association of Universities for Research
in Astronomy, Inc., under NASA contract NAS5-26555. Support for
MAST for non-HST data is provided by the NASA Office of Space
Science via grant NNX09AF08G and by other grants and contracts.

\end{document}